\documentclass[12pt]{article}
\usepackage{cite,epsfig,amssymb,amsmath,graphicx,color,subfigure}
\evensidemargin \oddsidemargin

\newcommand{\nn}{\nonumber}

\begin{document}

~
\vspace{4mm}
\begin{center}
{{\Large \bf Aspects of Massive ABJM Models \\ with Inhomogeneous Mass Parameters}}
\\[17mm]
Kyung Kiu Kim \\[2mm]
{\it  Department of Physics, Sejong University, Seoul 05006, Korea} \\
{\it kimkyungkiu@sejong.ac.kr}
\\[3mm]
Yoonbai Kim,~~ O-Kab Kwon$^*$
\\[2mm]
{\it Department of Physics,~BK21 Physics Research Division, Autonomous Institute of Natural Science$^*$, Institute of Basic Science, Sungkyunkwan University, Suwon 16419, Korea} \\
{\it yoonbai@skku.edu, ~okab@skku.edu}
\\[3mm]
Chanju Kim \\[2mm]
{\it Department of Physics, Ewha Womans University, Seoul 03760 Korea}
\\
{\it  cjkim@ewha.ac.kr}
\end{center}
\vspace{15mm}

\begin{abstract}
\noindent
Recently, ${\cal N} =3$ mass-deformed ABJM model 
with arbitrary mass-function depending on a spatial coordinate was constructed. 
In addition to the ${\cal N} = 3$ case, we construct lower supersymmetric ${\cal N} =1$ and ${\cal N} =2$ inhomogeneously mass-deformed ABJM (ImABJM) models, which require three and two arbitrary mass-functions, respectively. 
We also construct general vacuum solutions of the ${\cal N} = 3$ ImABJM model for any periodic mass-function. There are two classes of vacua,
which are diagonal type and GRVV type according to reference value of mass-functions. We provide explicit examples of the vacuum solutions and discuss related operators.
\end{abstract}

\newpage
 
\section{Introduction}

The low energy behavior of $N$ M2-branes on $\mathbb{Z}_k$-orbifold is 
described by the ${\cal N} = 6$ ABJM
theory with U$_k(N)\times {\rm U}_{-k}(N)$ gauge group at level $k$~\cite{Aharony:2008ug}. Due to the non-dynamical nature of the Chern-Simons gauge fields, an interesting extension is realized by use of  a supersymmetry-preserving mass deformation, which  gives rise to the mass-deformed ABJM (mABJM) theory~\cite{Hosomichi:2008jb,Gomis:2008vc}. The  mass parameter is originated from the self-dual 
constant 4-form field strength accompanied by the Wess-Zumino type coupling with M2 branes~\cite{Lambert:2009qw,Kim:2012gz} in the 11-dimensional supergravity. Gravity dual of the mABJM theory is identified  as the LLM geometry~\cite{Lin:2004nb} with $\mathbb{Z}_k$ orbifold. It also turns out that the 4-form field strength with one spatial coordinate dependence allows an ${\cal N} =  3$ 
supersymmetry  maximally. Accordingly, the ABJM theory with spatially dependent mass-function was constructed~\cite{Kim:2018qle}. We will call such a theory the inhomogeneously mass-deformed ABJM (ImABJM) model in what follows. The spatially varying mass-function $m= m(x)$ whose functional form is arbitrary breaks a half supersymmetry of the original mABJM theory of ${\cal N} = 6$ supersymmetry.

Gravity dual of the ImABJM model with periodic mass-functions is qualitatively similar to the holographic lattice models~\cite{Horowitz:2012ky}, which are dual to conformal field theories deformed by spatially periodic sources. The holographic lattice models in various gravity theories have been successful in describing many aspects including transport mechanism of condensed matter system with momentum relaxation {\it e.g.}~\cite{Vegh:2013sk,Andrade:2013gsa,Donos:2013eha,Blake:2013bqa,Davison:2013jba,Donos:2012js,Gouteraux:2014hca,Blake:2013owa,Donos:2014cya,Horowitz:2012gs,Davison:2014lua,Davison:2013txa,Horowitz:2013jaa,Baggioli:2014roa,Donos:2014yya,Ling:2013nxa,Donos:2014oha,Kim:2014bza,Ling:2014saa}. In particular, the Q-lattice solution~\cite{Donos:2013eha} was constructed to describe lattice structures with global symmetries in the bulk. One important advantage of  the Q-lattice is that the solution can be constructed by  solving ordinary differential equation rather than partial differential equation.  Later, supersymmetric  (SUSY) Q-lattice solution was found in 4-dimensional gauged supergravity, and then it can be uplifted to 11-dimensions~\cite{Gauntlett:2018vhk}. It was also generalized to the case of mass-functions with spatial modulations~\cite{Arav:2018njv}. For a special form of mass-function $m(x)\sim \sin (q\, x)$ with a constant $q$, it was argued that the corresponding dual field theory of the 11-dimensional Q-lattice solution is the ${\cal N} = 3$ ImABJM model \cite{Kim:2018qle} with the same mass-function. 
One of interesting properties of SUSY Q-lattices is that the geometries describe the boomerang RG flowing from the AdS$_4$ geometry in the UV to the same  AdS$_4$ geometry in the IR~\cite{Donos:2017ljs,Donos:2017sba,Chesler:2013qla}. 
On the other hand, there are other class of solutions~\cite{DHoker:2009lky,Bobev:2013yra} related to inhomogeneous mass functions in the 11-dimensional supergravity, which are not included in the Q-lattice solution. 
In particular, some of solutions in \cite{Bobev:2013yra} have lower supermetries with more mass functions, which may be related to the ${\cal N} = 1,2$ ImABJM models in this paper.

In \cite{Gauntlett:2018vhk,Arav:2018njv}, the strongly coupled limit of the vacuum expectation values (vevs) for chiral primary operators (CPOs) with the deformed ABJM Lagrangian $\Delta {\cal L} = m'(x) {\cal O}^1 + m(x) {\cal O}^2 + \cdots$ \cite{Kim:2018qle} was investigated by using the holographic method. 
It was shown that some special cases of the gravity description admit the SUSY Q-lattice solutions. For the weakly coupled limit of the ImABJM models, however, the holographic method is not a useful tool, while  the field theoretic perturbative methods  on vacuum solutions can be alternatives. In this point of view, studying vacuum structure of the ImABJM model itself would be an interesting subject to understand weak coupling behavior of the theory.

In this paper, we construct lower $({\cal N} =1,2)$ supersymmetric ImABJM models and  general vacuum solutions for the ${\cal N} = 3$ ImABJM model with periodic mass-functions. The $\mathcal{N}=3$ model has only one arbitrary mass-function, whereas ${\cal N} = 1$ and ${\cal N} = 2$ cases need two and three arbitrary mass-functions, respectively. We also show that there are two types of vacuum solutions according to the reference value of periodic mass-functions. Specifically, all vacuum configurations of scalar fields $Y_0^A$'s become diagonal, when $\int_0^\tau dx\, m(x)=0$ with a spatial period $\tau$. 
On the other hand, for $\int_0^\tau dx\, m(x)\ne 0$ case, the corresponding vacuum configurations $Y_0^A$'s are proportional to the GRVV 
matrices~\cite{Gomis:2008vc}, which construct vacuum solutions in the constant mass deformation of the ABJM theory. We also discuss conformal dimension $\Delta = 1,2$ CPOs in terms of classical limit of the vevs
and those gravity duals.

This paper is organized as follows. In section 2, we find a general supersymmetric condition for the space-dependent mass matrices and ${\cal N} = 6$ supersymmetric parameters. As special cases, we obtain ${\cal N} =1$ and ${\cal N} = 2$ supersymmetric ImABJM models. In section 3, we obtain conditions for supersymmetric vacuum energy configurations. We show that the energy of the vacuum configuration, which satisfies $\delta \psi_A=0$ for the variation of fermion fields, is determined by a boundary term only.
In section 4, we construct two types of general vacuum solutions depending on the shape of periodic mass-functions in the ${\cal N} = 3$ ImABJM model. In section 5, we conclude with a summary and a discussion for future directions. In Appendix A, we derive vacuum equation from the condition $\delta \psi_A=0$. In Appendix B, we display several vacuum solutions for the diagonal and GRVV types.

\section{${\cal N} =1, 2$ ImABJM Models}\label{N=3Janus}

The ABJM action with U$_k$($N)\times {\rm U}_{-k}(N$) gauge group at Chern-Simons level $k$ is given by \begin{align}\label{ABJMac}
S =\int d^3x\,{\cal L}_{{\rm ABJM}} = \int d^3x\,\left({\cal L}_0 + {\cal L}_{{\rm CS}} -V_{{\rm ferm}} -V_{{\rm bos}}\right),
\end{align}
 where 
 \begin{align}\label{Lagrangian00}
{\cal L}_0 &= {\rm tr}\left(-D_\mu Y_A^\dagger D^\mu Y^A +
i\psi^{\dagger A} \gamma^\mu D_\mu \psi_A\right),
\nn \\
{\cal L}_{{\rm CS}} &= \frac{k}{4\pi}\,\epsilon^{\mu\nu\rho}\,{\rm tr}
\left(A_\mu \partial_\nu A_\rho +\frac{2i}{3}A_\mu A_\nu A_\rho
- \hat{A}_\mu \partial_\nu \hat{A}_\rho
-\frac{2i}{3}\hat{A}_\mu \hat{A}_\nu \hat{A}_\rho \right),
\nn \\
V_{{\rm ferm}} &= \frac{2\pi i}k{\rm tr}\Big( Y_A^\dagger Y^A\psi^{\dagger
B}\psi_B -Y^A Y_A^\dagger\psi_B \psi^{\dagger B}
+2Y^AY_B^\dagger\psi_A\psi^{\dagger B} -2Y_A^\dagger
Y^B\psi^{\dagger A}\psi_B
\nn \\
&\hskip 1.7cm  +\epsilon^{ABCD}Y^\dagger_A\psi_BY^\dagger_C\psi_D
-\epsilon_{ABCD}Y^A\psi^{\dagger B}Y^C\psi^{\dagger D} \Big),
\nn \\
V_{{\rm bos}} &=-\frac{4\pi^2}{3k^2}{\rm tr}\Big(
Y^\dagger_AY^AY^\dagger_BY^BY^\dagger_CY^C
+Y^AY^\dagger_AY^BY^\dagger_BY^CY^\dagger_C
+4Y^\dagger_AY^BY^\dagger_CY^AY^\dagger_BY^C
\nn \\
&\hskip 2cm -6Y^AY^\dagger_BY^BY^\dagger_AY^CY^\dagger_C \Big).
\end{align} 
Here `tr' denotes the trace over gauge indices.
This ABJM action is invariant under the ${\cal N} = 6$ supersymmetric transformation,   $\delta_1 + \delta_2 +\delta_\mathcal{A}$,
\begin{align}\label{delta1}
&\delta _1 Y^A =i \omega ^{\text{AB}} \psi _B,
\qquad\qquad \delta _1 Y_A^{\dagger} = i \psi^{\dagger B}\omega_{AB}, 
\nn \\
&\delta_1\psi_A = \gamma^\mu \omega_{AB} D_\mu Y^B, 
\qquad 
\delta_1 \psi^{\dagger A} = -D_\mu Y_B^\dagger \omega^{AB}\gamma^\mu, 
\nn \\
&\delta_2 \psi_A= \frac{2\pi}{k} \omega_{AB}\left(Y^B Y_{C}^\dagger Y^C - Y^C Y_{C}^\dagger Y^B\right) + \frac{4\pi}{k} \omega_{BC} Y^B Y_A^\dagger Y^C,
\nn \\
&\delta_2 \psi^{\dagger A} = \frac{2\pi}{k} \omega^{AB}\left(Y_{C}^\dagger Y^C Y_B^\dagger- Y_{B}^\dagger Y^C Y_C^\dagger\right) - \frac{4\pi}{k} \omega^{BC} Y_B^\dagger Y^A Y_C^\dagger, 
\nn \\
&\delta_\mathcal{A} A_\mu =  -\frac{2\pi}{k}\left(Y^A\psi^{\dagger B}\gamma_\mu\omega_{AB} + \omega^{AB} Y_A^\dagger  \gamma_\mu\psi_B \right), 
\nn  \\
&\delta_\mathcal{A} \hat A_\mu = -\frac{2\pi}{k}\left(
\psi^{\dagger B} \gamma_\mu Y^A\omega_{AB} + \omega^{AB}Y_A^\dagger\gamma_\mu\psi_B \right),
\end{align}
where the supersymmetric parameters $\omega^{AB}$ satisfy the reality condition,  
\begin{align}\label{omegaAB}
\omega^{AB}=-\omega^{BA}=(\omega_{AB})^*=
\frac{1}{2}\,\epsilon^{ABCD}\omega_{CD}.
\end{align}

A noteworthy character of the ABJM theory is that it admits the supersymmetry preserving
mass deformation~\cite{Hosomichi:2008jb,Gomis:2008vc}. Assuming the mass parameter depends on one spatial coordinate $x$, the deformed action with this arbitrary mass-function $m(x)$ can preserve maximally ${\cal N} = 3$ supersymmetry~\cite{Kim:2018qle}, the so-called ImABJM model. In this section, we discuss general procedure extending to various ImABJM models preserving lower supersymmetry of  ${\cal N} =1,2,3$.

In order to construct the general ImABJM models, we begin with a deformation of the supersymmetry transformation rules for spinor fields,
\begin{align}\label{Jrule1}
\delta_J \psi_A &= \bar M_A^B(x) \omega_{BC} Y^C, 
\nn \\
\delta_J \psi^{\dagger A} &=\bar M^A_B(x) \omega^{BC} Y_C^\dagger,
\end{align}
where $\bar M_A^B(x)$'s are arbitrary space-dependent mass parameters. Application of the variation $\delta_J$ to the kinetic term ${\cal L}_0$ in \eqref{ABJMac} gives
\begin{align}\label{dJL0}
\delta_J {\cal L}_0 = \delta_1 \hat V_{{\rm ferm}} - i \left(\partial_\mu \bar M^A_B\right) \omega^{BC} {\rm tr}\big(Y_C^\dagger\gamma^\mu\psi_A\big) + i \left(\partial_\mu \bar M_A^B\right){\rm tr}\big(\psi^{\dagger A} \gamma^\mu\omega_{BC} Y^C\big),
\end{align}
where $\delta_1$ is defined in \eqref{delta1} and $\hat V_{{\rm ferm}}$ is given by
\begin{align}
\hat V_{{\rm ferm}} = i\bar  M_A^B{\rm tr}\big(\psi^{\dagger A}\psi_B\big).
\end{align}
Using the supersymmetric transformation rules \eqref{delta1} and the deformed one in  \eqref{dJL0}, we obtain 
\begin{align}\label{sr2}
&\delta_{{\rm tot}}\left({\cal L}_{{\rm ABJM}} - \hat V_{{\rm ferm}}  \right)+ \delta_J V_{{\rm ferm}} + (\delta_2 +\delta_J)\hat V_{{\rm ferm}} \nn \\
&= - i \left(\partial_\mu \bar M^A_B\right) \omega^{BC} {\rm tr}\big(Y_C^\dagger\gamma^\mu\psi_A\big) + i \left(\partial_\mu \bar M_A^B\right){\rm tr}\big(\psi^{\dagger A} \gamma^\mu\omega_{BC} Y^C\big),
\end{align}
where  $\delta_{{\rm tot}} = \delta_1 + \delta_2 + \delta_\mathcal{A} + \delta_J$.
In order to complete the supersymmetry transformation of the deformed ABJM theory in \eqref{sr2}, we perform the transformation $\delta_J$ on the mass deformation $ V_{{\rm ferm}}$  in \eqref{ABJMac} and consequently it turns out that 
\begin{align}\label{dJVf}
\delta_J V_{{\rm ferm}} &= i \bar \mu_B^E {\rm tr}\big(\psi^{\dagger B} \omega_{EF} \beta^{FA}_A\big) + 2i\bar  M_A^E{\rm tr}\big(\psi^{\dagger B} \omega_{EF} \beta_B^{AF} \big)
\nn \\
&~~~- \frac{i}2 \bar M_E^B \epsilon_{ABCD}\epsilon^{EFGH}{\rm tr}\big(\psi^{\dagger D}\omega_{GH} \beta_F^{AC}\big)+{\rm (c.c.)}
\nn \\
&= - \delta_2 \hat V_{{\rm ferm}} - \delta_{tot} \hat V_{{\rm flux}},
\end{align}
where the (c.c.) denotes the complex conjugate of the previous expressions and the $\beta$ matrices are defined by
\begin{align}\label{beta}
&\beta^{AB}_{\; C}
  \equiv \frac{2\pi}{k}\left(Y^{A}Y_{C}^{\dagger}Y^{B}-Y^{B}Y_{C}^{\dagger}Y^{A}\right),
\nn \\
&\beta^{A}_{BC}
  \equiv \frac{2\pi}{k}\left(Y^{\dagger}_BY^AY_{C}^\dagger-Y^{\dagger}_CY^AY_{B}^\dagger\right).
\end{align}
Also, the quartic scalar term $\hat V_{{\rm flux}}$ in \eqref{dJVf} is given by 
\begin{align}\label{VVff}
\hat V_{{\rm flux}} &= -2\bar  M_A^B {\rm tr}\big(Y_B^\dagger \beta_C^{AC}\big).
\end{align}
In the last equality of \eqref{dJVf}, we imposed the traceless condition of the mass matrix $\bar M_A^B$, 
\begin{align}\label{mAA}
\bar M_A^A =0.
\end{align}
Plugging \eqref{dJVf} into \eqref{sr2}, we obtain 
\begin{align}\label{sr3}
&\delta_{{\rm tot}}\left({\cal L}_{{\rm ABJM}}  - \hat V_{{\rm ferm}} -\hat V_{{\rm flux}}\right)+ \delta_J \hat V_{{\rm ferm}} \nn \\
&= - i \left(\partial_\mu \bar M^A_B\right) \omega^{BC} {\rm tr}\big(Y_C^\dagger\gamma^\mu\psi_A\big) + i \left(\partial_\mu \bar M_A^B\right){\rm tr}\big(\psi^{\dagger A} \gamma^\mu\omega_{BC} Y^C\big).
\end{align}
We also have the expression for action of $\delta_J$ to $\hat V_{{\rm ferm}}$:
\begin{align}
\delta_J \hat V_{{\rm ferm}}
= -\bar  \mu_{A}^B \delta_{\rm tot} {\rm tr}(Y_B^\dagger Y^A),
\end{align} 
where $\bar \mu_A^B$ is  another mass matrix satisfying the relations with the mass matrix $\bar M_A^B$ as
\begin{align}\label{mmom}
\bar M_A^B \bar M_B^C \omega_{CD} - \bar \mu_D^B \omega_{AB} = 0.
\end{align}

For the case of the constant mass parameter of $\partial_\mu \bar M_A^B = 0$, the right-hand side of \eqref{sr3} vanishes and thus the two conditions \eqref{mAA} and \eqref{mmom} are automatically satisfied for a particular  choice of mass matrices, 
\begin{align}\label{N=6mABJM}
&\bar M_A^B = {\rm diag} (m,m,-m,-m),
\nn \\
&\bar \mu_A^B =  {\rm diag}(m^2,m^2,m^2,m^2)
\end{align}
without any restriction on $\omega_{AB}$. In synthesis, the total Lagrangian, 
\begin{align}\label{calLm}
{\cal L}_{m} = {\cal L}_{{\rm ABJM}} - \hat V_{{\rm ferm}}  -\hat V_{{\rm flux}}- \hat V_{{\rm mass}},
\end{align}
including a quadratic scalar mass term,
\begin{align}\label{Vmass}
\hat V_{{\rm mass}} = \bar \mu_{A}^B  {\rm tr}\big(Y_B^\dagger Y^A\big),
\end{align} 
preserves the ${\cal N} = 6$ supersymmetry of the original ABJM theory~\cite{Hosomichi:2008jb,Gomis:2008vc}. 
In what follows, we take into account  the   spatially varying mass parameters of $\partial_\mu \bar M_A^B \ne 0$, and check the amount of residual supersymmetries.

\subsection{${\cal N} = 3$ deformation}\label{N=3def}

Suppose that the mass parameter $m$ in \eqref{N=6mABJM} becomes a mass-function $m(x)$ depending on a spatial coordinate $x$ in the mABJM theory in \eqref{calLm}. An intriguing question at the moment is about the residual supersymmetry from the original ${\cal N} =6$ supersymmetry. The right-hand side (RHS) of \eqref{sr3} is given by 
\begin{align}
{\rm RHS} ={\rm tr}\Big(-i m' \bar M^A_{~B} Y_C^\dagger \omega^{BC} \gamma^1 \psi_A + i m' \bar M_{A}^{~B}\psi^{\dagger A}\gamma^1 \omega_{BC} Y^C\Big),
\end{align}
where $m'(x) = \frac{dm}{dx}$.  
Under the following projection of the supersymmetric parameter by  $\gamma^1$, 
\begin{align}\label{proj}
\gamma^1 \omega_{ab}=-\omega_{ab}\quad &\longleftrightarrow\quad \omega^{ab}\gamma^1 =\omega^{ab}, 
\nn \\
\gamma^1 \omega_{ai}=\omega_{ai}\quad &\longleftrightarrow\quad \omega^{ai}\gamma^1 =-\omega^{ai},
\end{align}
 we obtain \begin{align}\label{approj}
{\rm RHS} &={\rm tr}\left[ \,i m'\left( Y_b^\dagger\omega^{ba}\psi_a - Y_i^\dagger\omega^{ia}\psi_a +Y_a^\dagger\omega^{ai}\psi_i - Y_j^\dagger\omega^{ji}\psi_i\right) \right]+ ({\rm c.c.})
\nn \\
&= {\rm tr}\left[m' \delta_{{\rm tot}}\left(Y_a^\dagger Y^a- Y_i^\dagger  Y^i\right)\right].
\end{align}
Inserting \eqref{approj} into \eqref{sr3} with the help of the projection \eqref{proj}, we obtain the ${\cal N} = 3$  
ImABJM model~\cite{Kim:2018qle},
\begin{align}\nn
\delta_{{\rm tot}}\left({\cal L}_{{\rm ABJM}} - \hat V_{{\rm ferm}}   -\hat V_{{\rm flux}}- \hat V_{{\rm mass}}-\hat V_J\right)=0,
\end{align}
where the quadratic mass term is
\begin{align} \label{VJ}
\hat V_J = m' {\rm tr}\left(Y_a^\dagger Y^a- Y_i^\dagger  Y^i\right).
\end{align}

\subsection{${\cal N} = 2$  deformation}
The ${\cal N} =3$ ImABJM model is derived from the ${\cal N} = 6$ mABJM theory in the previous subsection. Naturally enough, we begin with the ${\cal N} = 4$ supersymmetric constant mass deformation of the ABJM theory~\cite{Kim:2012gz} and construct the ${\cal N} =2$ Inhomogeneous deformation in this subsection.  
With specific mass matrices,
\begin{align}\label{N=2mm}
&\bar M_A^B = {\rm diag} (m_1,-m_1,m_2,-m_2) ,
\nn \\
&\bar \mu_A^B = {\rm diag} (m_2^2,m_2^2,m_1^2,m_1^2),
\end{align}
the supersymmetric conditions  \eqref{mAA} and \eqref{mmom} are satisfied with two nonvanishing supersymmetric parameters,
\begin{align}
\omega_{12} =0,\quad \omega_{13}\ne 0\,\,\,\&\,\,\, \omega_{14} \ne 0, 
\end{align}
where no additional restriction is assigned to the nonvanishing parameters $\omega_{13}$ and $\omega_{14}$. Therefore, the resultant deformed theory possesses ${\cal N} = 4$ supersymmetry for constant mass parameters of $\partial_\mu \bar M_A^B= 0$, which make the right-hand side of \eqref{sr3} vanish. 

For  mass-functions $m_i= m_i(x)$ varying along $x$, the right-hand side of \eqref{sr3} is written as
\begin{align}
{\rm RHS} &={\rm tr}\big(-i m_1'\psi^{\dagger 1} \gamma^1 \omega_{31} Y^3-i m_1'\psi^{\dagger 1} \gamma^1 \omega_{41} Y^4
+i m_1'\psi^{\dagger 2} \gamma^1 \omega_{32} Y^3+i m_1'\psi^{\dagger 2} \gamma^1 \omega_{42} Y^4
\nn \\
&~~-i m_2'\psi^{\dagger 3} \gamma^1 \omega_{13} Y^1-i m_2'\psi^{\dagger 3} \gamma^1 \omega_{23} Y^2
+i m_2'\psi^{\dagger 4} \gamma^1 \omega_{14} Y^1+i m_2'\psi^{\dagger 4} \gamma^1 \omega_{24} Y^2\big).
\end{align}
Imposing the projection conditions,
\begin{align}\label{gproj2}
\gamma^1 \omega_{13} = \omega_{13},\quad \gamma^1 \omega_{14} = - \omega_{14},
\end{align}
one can rewrite the ${\rm RHS}$ as 
\begin{align}\label{RHS2}
{\rm RHS} ={\rm tr}\big[ -m_1'\delta_{{\rm tot}}(Y_3^\dagger  Y^3)+m_1'\delta_{{\rm tot}}(Y_4^\dagger  Y^4)-m_2'\delta_{{\rm tot}}(Y_1^\dagger  Y^1)+m_2'\delta_{{\rm tot}}(Y_2^\dagger  Y^2)\big].
\end{align}
Insertion of \eqref{RHS2} into \eqref{sr3} gives 
\begin{align}
\delta_{{\rm tot}} {\cal L}_{{\cal N} =2} =0,
\end{align}
where the terms are
\begin{align}\label{N=2Lag}
&{\cal L}_{{\cal N} =2} = {\cal L}_{{\rm ABJM}} - \hat V_{{\rm ferm}}   -\hat V_{{\rm flux}}- \hat V_{{\rm mass}}-\hat V_J,
\nn \\
&\hat V_J = {\rm tr}\big[m_1'(Y_3^\dagger  Y^3-Y_4^\dagger  Y^4)+m_2'(Y_1^\dagger  Y^1-Y_2^\dagger  Y^2)\big] .
\end{align}
Due to the projections in \eqref{gproj2}, the  supersymmetry  of the current ImABJM model becomes ${\cal N} = 2$, which includes two arbitrary mass-functions.

\subsection{${\cal N} = 1$  deformation }
Above all we recall the ${\cal N} = 2$ mass-deformed ABJM theory with a constant mass parameter~\cite{Kim:2012gz}.
When the following mass matrices,
\begin{align}\label{N=1mm}
&\bar M_A^B = {\rm diag} (m_1,m_2,m_3,m_4) \,\,\, {\rm with\,\, a\,\, constraint}\,\, m_1 + m_2 + m_3 +m_4= 0,
\nn \\
&\bar \mu_A^B = {\rm diag} (m_2^2, m_1^2, m_4^2,m_3^2),
\end{align}
are chosen, the conditions \eqref{mAA} and \eqref{mmom} are satisfied only under the conditions of supersymmetric parameters,
\begin{align}
\omega_{12} \ne 0,\quad \omega_{13} =\omega_{14} =0, 
\end{align}
 where the nonvanishing  parameter $\omega_{12}$ has no further restriction. For the constant mass parameters \eqref{N=1mm},  the ${\cal N} = 2$ supersymmetry is preserved in this mass-deformed ABJM theory, as shown in \eqref{sr3}. Lower supersymmetric mass deformations and their RG flows to IR limit were also discussed in \cite{Bobev:2009ms}.

Now, we turn on coordinate dependence to the  mass parameters $m_i$'s. Then the right-hand side of \eqref{sr3} becomes 
\begin{align}
{\rm RHS} 
&={\rm tr}\big(i m_1'\psi^{\dagger 1} \gamma^1 \omega_{12} Y^2+ im_2'\psi^{\dagger 2} \gamma^1 \omega_{21} Y^1
\nn \\
&~~~~~~~+im_3'\psi^{\dagger 3} \gamma^1 \omega_{34} Y^4 +im_4'\psi^{\dagger 4} \gamma^1 \omega_{43} Y^3 \big)+({\rm c.c.})
\end{align}
with a constraint $m_1' + m_2' + m_3' +m_4'= 0$. Under the following projection of supersymmetric parameters: 
\begin{align}\label{N=1proj}
\gamma^1\omega_{12} = -\omega_{12},\quad \gamma^1\omega_{34} = -\omega_{34},
\end{align}
we obtain 
\begin{align}\label{RHS1}
{\rm RHS} ={\rm tr}\big[ m_1'\delta (Y_2^\dagger Y^2 )+m_2'\delta (Y_1^\dagger Y^1 )+m_3'\delta (Y_4^\dagger Y^4 )+m_4'\delta (Y_3^\dagger Y^3 )\big].
\end{align}
Insertion of \eqref{RHS1} into \eqref{sr3} results in 
\begin{align}
\delta_{{\rm tot}} {\cal L}_{{\cal N} =1} =0,
\end{align}
where the Lagrangian is given by
\begin{align}\label{N=1Lag}
{\cal L}_{{\cal N} =1} = {\cal L}_{{\rm ABJM}} - \hat V_{{\rm ferm}}   -\hat V_{{\rm flux}}- \hat V_{{\rm mass}}-\hat V_J.
\end{align}
The second and third terms are in \eqref{VVff}, the 
fourth term is in \eqref{Vmass}, and the last term is 
\begin{align}
\hat V_J ={\rm tr}\big[ m_1' (Y_2^\dagger Y^2-Y_3^\dagger Y^3) +m_2' (Y_1^\dagger Y^1 -Y_3^\dagger Y^3)+m_3' (Y_4^\dagger Y^4-Y_3^\dagger Y^3)\big] .
\end{align}
Due to the projection for the nonvanishing supersymmetric parameters in \eqref{N=1proj}, the deformed Lagrangian in \eqref{N=1Lag} preserves the ${\cal N} = 1$ supersymmetry involving three arbitrary mass-functions.

In the remaining part of this paper, we focus on the  ${\cal N} = 3$ ImABJM model and investigate some aspects of it, such as discussion on gravity duals and vacuum solutions for some specific mass-functions.

\section{Primary Conditions for Vacuum Solutions }

Recently, gravity duals to the ImABJM theory  with spatially modulated mass-functions have been studied using the AdS/CFT correspondence~\cite{Gauntlett:2018vhk,Arav:2018njv}. This approach allows to find various physical quantities in the strongly coupled ImABJM theory with the large $N$ limit. However, the weak-coupling limit is also interesting to understand vacuum structures of the theory and to compare them with those in the strong coupling limit. Therefore we try to obtain general vacuum equations and boundary conditions for supersymmetric vacua in the $\mathcal{N}=3$ ImABJM model. As a first attempt, we consider supersymmetric configuration without central extension. Assuming there is no central charge, the supersymmetric configurations satisfy the trivial energy condition, i.e, $E_{vac}=0$. In this paper we focus on this case\footnote{Since more careful study is needed for nontrivial vacua with nonvanishing central charges, which give rise to negative energy solutions with $E_{vac}<0$. We will clarify this issue in our upcoming project \cite{KKKK}.}. For asymptotically constant mass-functions, we discuss the boundary condition for vacuum solutions connecting two different vacua denoted by GRVV matrices~\cite{Gomis:2008vc} in the mABJM theory.\footnote{These matrix solutions were also obtained in \cite{Terashima:2008sy} to describe the M2-M5 bound state in the ABJM theory.}  The periodic configurations will be taken into account in the next section.

\subsection{Deformed operators and energy-momentum tensors}
As we explained in the subsection \ref{N=3def}, the $\mathcal{N}=3$ ImABJM theory can be obtained by the following deformation coming from inhomogeneous mass sources:
\begin{align}\label{delL}
\Delta \mathcal{L} =  m'(x)\, \mathcal{O}^{ 1} + m(x)\, \mathcal{O}^{2} + m(x)^2\, \mathcal{O}^{3}~,
\end{align} 
where 
\begin{align}
&\mathcal{O}^1\equiv \mathcal{O}^{\Delta=1} = M_A^B {\rm tr} \left(Y^A Y_B^\dagger\right), 
\nn \\
&\mathcal{O}^2\equiv\mathcal{O}^{\Delta=2} =M_A^B {\rm tr}\left( \psi^{\dagger A} \psi_B + \frac{8\pi}{k} Y^C Y_{[C}^\dagger Y^A Y_{B]}^\dagger\right), \nn \\
&\mathcal{O}^3\equiv\mathcal{O}^K =  {\rm tr}\left( Y^A Y_A^\dagger \right).\label{Operators}
\end{align}
Here $M_A^B = {\rm diag}(1,1,-1,-1)$ and the first two operators are chiral primary operators. Also the third operator $\mathcal{O}^K$ is the Konishi operator, which is not protected by supersymmetry and large $N$ limit. By this reason, this operator does not play an important role in the dual gravity solution. But it is still worth considering in the weak coupling limit. Thus we will provide the classical vevs of these operators using the general vacuum solutions in the next section. Some of them may be compared to the same quantities in the strong coupling limit.

Another important quantity in this system is the energy-momentum tensor. Since the fermionic part does not have the vacuum expectation value, we write down only the bosonic part of the energy-momentum tensor as follows:
\begin{align} \label{Tmn1}
T^{\mu\nu} = {\rm tr} \left(D^\mu Y_A^\dagger D^\nu Y^A + D^\nu Y_A^\dagger D^\mu Y^A\right) - \eta^{\mu\nu}\left[{\rm tr}\left(D^\rho Y_A^\dagger D_\rho Y^A \right)+ \hat V_{{\rm bos}} \right],
\end{align}
where
\begin{align}\label{Vhbos0}
&\hat V_{{\rm bos}} = V_{\rm bos} +\hat  V_{{\rm flux}} +\hat V_{{\rm mass}} +\hat V_J.
\end{align} 
The $V_{{\rm bos}}$ is defined 
in \eqref{ABJMac} and the other deformed potentials for the ${\cal N} = 3$ case are given by 
\begin{align}
&\hat V_{{\rm flux}} = -2m\,M_A^B {\rm tr}\big(Y_C^\dagger\, \beta_B^{A\,C} \big),\nn \\
&\hat V_{{\rm mass}} = m^2 {\rm tr}\big(Y_A^\dagger Y^A\big),
\nn \\
&\hat V_J = m' M_A^{~B}{\rm tr}\big( Y_B^\dagger Y^A\big).
\end{align}

\subsection{Vacuum energy}

Due to the inhomogeneity, the energy density can be a spatially dependent function. However, the total energy should vanishes to preserve supersymmetry. 
In order to construct the vacuum equations and the corresponding boundary conditions, we read the total energy of the bosonic sector from the energy-momentum tensor in \eqref{Tmn1}, \begin{align}\label{energy1}
E = \int d^2 x\left[{\rm tr}\left( |D_0 Y^A|^2 + |D_i Y^A|^2\right)+ \hat V_{{\rm bos}}\right],
\end{align} 
where the potential $\hat V_{{\rm bos}}$ in (\ref{Vhbos0}) can be rearranged 
in the following form~\cite{Kim:2009ny},
\begin{align}\label{Vhbos}
\hat V_{{\rm bos}} = \frac23 {\rm tr} \left|G^{BC}_A\right|^2 + \hat V_J
\end{align}
with 
\begin{align}\label{GABC}
G^{BC}_A = \beta^{BC}_A + \delta^{[ B}_{A} \beta^{C]D}_D + m M_A^{~[B} Y^{C]}.
\end{align}
Here we use the notation $|{\cal O}|^2 \equiv {\cal O}^\dagger {\cal O}$
for simplicity. 
For the constant mass parameter($\partial_\mu m=0$), the vacuum configuration in terms of GRVV matrices~\cite{Gomis:2008vc} are given by homogeneous nonvanishing scalar fields together with vanishing fermion and gauge fields. Once an inhomogeneous mass deformation is turned on along the $x$-direction, the vacuum configurations with the scalar fields also depend on the  $x$-direction. To reflect this, we rewrite the energy in \eqref{energy1} as 
\begin{align}\label{energy2}
E_{{\rm vac}} = \int d^2 x\left({\rm tr} |Y^{A'}|^2 + 
\hat V_{{\rm bos}}\right)= \int d^2 x\left[{\rm tr}\left( |Y^{A'}|^2 + \frac23|G^{BC}_A|^2\right) + \hat V_J\right].
\end{align}

Recall that the supersymmetry variation of fermion fields should vanish for a  supersymmetric vacuum, i.e.,
\begin{align}\label{psiA0}
\delta_{{\rm tot}}\psi_A = 0.
\end{align}  
Since we are interested in vacuum configurations, the gauge fields vanish and the scalar fields depend on the $x$-coordinate, $Y^A = Y^A(x)$, \begin{align}\label{fertr}
\delta_{{\rm tot}} \psi_A
&=\gamma^\mu \omega_{AB} D_\mu Y^B + \omega_{BC} G^{BC}_A
\nn \\
&=\left(\delta^{[B}_A Y^{C]'}s_{BC} + G^{BC}_A\right) \omega_{BC}.
\end{align}

In the above \eqref{fertr}, a useful notation is employed for describing the projection (\ref{proj}) for the ${\cal N} = 3$ ImABJM theory, 
\begin{align}
\gamma^1\omega_{AB} = s_{AB}\, \omega_{AB,}\qquad (s_{ab} =s_{ij}= -1,\, s_{ai} = s_{ia} =1),
\end{align} 
where $a,b = 1,2$ and $i,j = 3,4$, and the repeated indices in the right-hand side are not summed. Manipulating the absolute square of the bosonic part in \eqref{fertr}, we obtain the following expression including a total derivative term: 
\begin{align}\label{absrel}
&|\delta^{[B}_A Y^{C]'} s_{BC} + G^{BC}_A|^2
\nn \\
&= \frac{3}{2}\left[|Y^{A'}|^2 +\hat V_{{\rm bos}}-\Big(\frac12\beta^{ac}_a Y_c^{\dagger}  + \frac12\beta^{ik}_i Y_k^{\dagger} -\frac13  \beta^{ic}_i Y_c^\dagger
+ m M_A^{~B} Y_B^\dagger Y^A\Big)'\right].
\end{align}
Detailed derivation of \eqref{absrel} is given in  Appendix \ref{urels}.
Together with the relation \eqref{absrel}, the energy \eqref{energy2} for the vacuum configuration becomes  
\begin{align}\label{Evac1}
E_{{\rm vac}}
&= \int d^2 x\,\left( \frac{2}{3}{\rm tr}\big|\delta^{[B}_A Y^{C]'} s_{BC}+ G^{BC}_A\big|^2+{\cal K} '\right),
\end{align}
where the total derivative term is
\begin{align}\label{calK}
\mathcal{K} \equiv \text{tr}\left(   \frac12\beta^{ac}_a Y_c^{\dagger}  + \frac12\beta^{ik}_i Y_k^{\dagger} -\frac13  \beta^{ic}_i Y_c^\dagger
+ m M_A^{~B} Y_B^\dagger Y^A \right).
\end{align}

\subsection{Conditions for supersymmetric vacua}

The Hamiltonian $H$ is represented as the anti-commutator of the supercharges in the supersymmetry algebra without central charges. Correspondingly, the vacuum expectation value of Hamiltonian should vanish for any supersymmetric vacuum, 
\begin{align}\label{Evac2}
E_{{\rm vac}} = \langle H \rangle = 0.
\end{align}
 Furthermore, the supersymmetry variation of the fermion fields also vanishes, as discussed in \eqref{psiA0}. Combining \eqref{fertr} and \eqref{Evac1}, the supersymmetric condition \eqref{psiA0} and \eqref{Evac2} leads to the  two relations, 
\begin{align}\label{vcon1}
&(i)\,\, \delta^{[B}_A Y^{C]'} +s^{BC} G^{BC}_A= 0, \\
&(ii) \int_{x_L}^{x_R} dx\, \mathcal{K}'=0,\label{vcon2}
\end{align}
where $x_L$ and $x_R$ denote the asymptotic boundaries  along the $x$-direction.
From now on, we call the first relation \eqref{vcon1} the vacuum equation, which will be analyzed in details in the next section. In addition the second relation \eqref{vcon2} is regarded as a constraint on the boundary condition for the $x$-direction. 

There are two types of representative inhomogeneous mass-functions. The first type consists of periodic mass-functions, whose gravity dual was already studied in \cite{Gauntlett:2018vhk,Arav:2018njv}. In this type, \eqref{vcon2} is trivially satisfied since one may take $\left(x_R-x_L\right)$ as an integer times the period of the mass-functions. So only the vacuum equation (\ref{vcon1}) remains to get the supesymmetric vacua. On the other hand, the second type we consider is the class of mass-functions which are asymptotically constant, i.e., 
\begin{align}\label{mLR}
\lim_{x_L \to -\infty}m(x_L) = m_L,\qquad \lim_{x_R\to\infty}m(x_R) = m_R,
\end{align}
where $m_L$ and $m_R$ are arbitrary constants.
As we know, there are many Higgs vacua denoted by GRVV matrices\cite{Gomis:2008vc} in the mABJM theory. This implies the existence of vacuum solutions, which connect a supersymmetric vacuum~\cite{Kim:2010mr} parametrized by $m_L$ at $x\to-\infty$ and another vacuum parametrized by $m_R$ at $x\to\infty$, if there are no central charges. However, in the case that the values of ${\cal K}$ \eqref{calK} for two vacua are different,  i.e., $\mathcal{K}_{x=-\infty} \neq \mathcal{K}_{x=\infty}$, the condition \eqref{vcon2} is not satisfied and then the corresponding configurations cannot be vacuum solutions. In order to construct vacuum solutions of the second type, finding supersymmetric Higgs vacua satisfying the condition \eqref{vcon2} is an important primary check.

Now, let us find general boundary conditions for the case of asymptotically constant mass-functions. As we will see the vacuum equations for the ${\cal N} =3$ ImABJM theory in \eqref{veq12}--\eqref{veq14} in section 4, the vacuum equations for a constant mass parameter are given by 
\begin{align}\label{cme1}
& \beta^{ca}_{c}+ m Y^a=0,
\nn \\
&\beta^{ki}_{k}- m Y^i = 0,
\nn \\
&\beta^{ja}_{i} =  \beta^{bi}_{a} =\beta^{ab}_{i}=\beta^{ij}_{a}=0.
\end{align}
With the help of \eqref{cme1}, the quantity $\mathcal{K}$'s defined in \eqref{calK} have the following asymptotic values in terms of \eqref{mLR}:
\begin{align}\label{cocalK}
{\cal K}|_{x\to x_i} = \frac{m_i}{2}{\rm tr} \left(  M_A^{~B} Y_B^\dagger Y^A\right)_{{\rm vac}},
\end{align}
where $i = L,R$ with $x_L = -\infty$ and $x_R = \infty$. To satisfy the condition \eqref{vcon2}, the values of ${\cal K}$'s for both asymptotic regions should be identified as 
\begin{align}\label{calK3}
{\cal K}|_{x\to-\infty} = {\cal K}|_{x\to \infty}.
\end{align}  
The quantity ${\rm tr} \left(  M_A^{~B} Y_B^\dagger Y^A\right)_{{\rm vac}}$ in the right-hand-side of \eqref{cocalK} was calculated~\cite{Jang:2016aug,Jang:2017gwd} for all possible supersymmetric vacua for given $k$ and $N$ in the mABJM theory and it was generalized to the mABJ theory with discrete 
torsions~\cite{Jang:2019pve}. To write down the quantity ${\cal K}|_{x\to x_i}$, we briefly review the supersymmetric vacua of the mABJM theory.

The general solutions of the vacuum equations in \eqref{cme1} were constructed using GRVV matrices~\cite{Gomis:2008vc}. For given $k$ and $N$, there are many possible vacuum solutions represented by direct sums of two types of irreducible $n\times (n+1)$ GRVV matirces ${\cal M}_\alpha^{(n)}$ $(\alpha=1,2)$ and their Hermitian conjugates, $\bar {\cal M}_\alpha^{(n)}$,
\begin{align}\label{GRVV1}
\mathcal{M}_1^{(n)} = \left(
\begin{matrix}
\sqrt{n}&0& & & &\\
 & \sqrt{n-1}&0 & & & \\
 & &\ddots &\ddots & & \\
 & & &\sqrt{2} &0 & \\
 & & & & 1& 0
\end{matrix}
\right),\,~ \mathcal{M}_2^{(n)} = \left(
\begin{matrix}
0&1& & & &\\
 &0&\sqrt{2} & & & \\
 & &\ddots &\ddots & & \\
 & & &0 &\sqrt{n-1} & \\
 & & & & 0& \sqrt{n}
\end{matrix}
\right),
\end{align}
where $n = 0,1,\cdots, N-1$.
Then the vacuum solutions satisfying \eqref{cme1} are given by 
\begin{align}\label{mABJMv}
Y^a|_{{\rm vac}}&=\sqrt{\frac{m k}{2\pi}}\left(\begin{array}{c}
\begin{array}{cccccc}\mathcal{M}_a^{(n_1)}\!\!&&&&&\\&\!\!\ddots\!&&&&\\
&&\!\!\mathcal{M}_a^{(n_i)}&&& \\ &&& {\bf 0}_{(n_{i+1}+1)\times n_{i+1}}
&&\\&&&&\ddots&\\&&&&&{\bf 0}_{(n_f+1)\times n_f}\end{array}\\
\end{array}\right),\nonumber
\end{align}
\begin{align}
Y^i|_{{\rm vac}}&=\sqrt{\frac{m k}{2\pi}}\left(\begin{array}{c}
\begin{array}{cccccc}{\bf 0}_{n_1\times (n_1+1)}&&&&&\\&\ddots&&&&\\
&&{\bf 0}_{n_i\times(n_i + 1)} &&&\\
&&& \bar{\mathcal M}_a^{(n_{i+1})}\!\!&&\\&&&&\!\!\ddots\!&\\
&&&&&\!\!\bar{\mathcal M}_a^{(n_f)}\end{array}\\
\end{array}\right).
\end{align}
The solution contains $N_n$ number of the type ${\cal M}_\alpha^{(n)}$ and $N_n'$ number of  the type $\bar{\cal M}_\alpha^{(n)}$, where $N_n$ and $N'_n$ are referred as occupation 
numbers~\cite{Kim:2010mr,Cheon:2011gv}. These numbers, $N_n$ and $N_n$, classify all possible supersymmetric vacua in the mABJM theory. These vacua have one-to-one correspondence with the LLM geometry~\cite{Lin:2004nb} with $\mathbb{Z}_k$ orbifold in 11-dimensional supergravity.

For a given supersymmetric vacuum with a set of occupation numbers $\{N_n,N_n'\}$,  we have \cite{Jang:2016aug,Jang:2017gwd,Jang:2019pve},
\begin{align}\label{MYYd}
{\rm tr} \left(  M_A^{~B} Y_B^\dagger Y^A\right)_{{\rm vac}}=\frac{m k}{2\pi}\sum_{n=0}^{\infty}
\big[n(n+1)(N_n-N_n')\big].
\end{align}
In general, vacuum solutions of the ${\cal N} =3$ ImABJM theory with asymptotically constant masses, $m_L$ and $m_R$, can connect different supersymmetric vacua with occupation numbers, $\{N_n^{(L)}, {N'}_n^{(L)}\}$ and $\{N_n^{(R)}, {N'}_n^{(R)}\}$, respectively. In this case, the boundary condition \eqref{vcon2} can be rewritten in terms of \eqref{MYYd} as 
\begin{align}
m_L^2\sum_{n=0}^\infty \big[n(n+1)(N^{(L)}_n-{N'}_n^{(L)})\big]= m_R^2\sum_{n=0}^\infty \big[n(n+1)(N^{(R)}_n-{N'}_n^{(R)})\big].
\end{align}
For $N_n = N'_n$ cases known as the symmetric vacuum solutions of the ${\cal N} =6$ mABJM theory, the ${\cal K}$'s in asymptotic limits are vanishing. Therefore, if there is no central charge, vacuum solutions in the ${\cal N} = 3$ ImABJM theory, which connect symmetric vacua  of the mABJM theory, trivially satisfy the constraint \eqref{vcon2}. For general $N_n$ and $N_n'$, one need more investigations.

\section{Periodic Vacuum Solutions for ${\cal N} =3$ Deformation}\label{pvacsol}

In this section, we consider the   ${\cal N} =3$ deformation again and investigate the general vacuum structure for  spatially modulated mass-functions. To do that, we try to solve the vacuum equation \eqref{vcon1}. More explicit form of the equation is as follows:    
\begin{align}
&Y^{a'}-  \beta^{ca}_{c}- m Y^a = 0,
\label{veq12} \\
&Y^{i'}-  \beta^{ki}_{k}+ m Y^i = 0,
\label{veq13} \\
&\beta^{ja}_{i} =  \beta^{bi}_{a} =\beta^{ab}_{i}=\beta^{ij}_{a}=0, \quad {\rm for\, any\,}\, a,b=1,2\,~{\rm and}\, i,j= 3,4, 
\label{veq14}
\end{align}
where $m(x)$ is a periodic function. Due to the periodic behavior, the boundary
term \eqref{vcon2} vanishes always and hence the energy of the solution
satisfying \eqref{vcon1} is guaranteed to be zero. So after solving (\ref{veq12}) and (\ref{veq13}), there is only one thing to check if the solutions is regular everywhere.

\subsection{$ m(x) = m_1 \sin{qx}$ case}\label{mx=m1s}
Since the solution space could depend on the mass-function, we would like to discuss
some representative examples. Let us start with a mass-function,
\begin{equation} \label{sinqx}
m(x) = m_1 \sin{qx},
\end{equation}
where $m_1 > 0$ and $q= 2\pi n /(x_R - x_L)$ for some integer $n$. For simplicity, we find a solution with $Y^a=0$. We will show that this example can be regarded as a periodic modulation of the (massless) ABJM theory.
It is useful to define a new variable $\xi(x)$ by
\begin{align} \label{xi}
&\xi(x) = \int^x e^{-2 \int^{x'} m(x'')\,dx''} dx' =c \int^x e^{\frac{2m_1}q \cos qx'} dx',
\end{align}
where $c$ is an integration constant.
Then using
\begin{align} \label{tildey}
 Y^i = e^{- \int m\, dx}\, \tilde Y^i(x),
\end{align}
\eqref{veq13} reduces to%
\begin{equation} \label{tildeyeq}
\frac{d}{d\xi} \tilde Y^i - \tilde\beta^{ki}_{k} = 0,
\end{equation}
where $\tilde\beta^{ij}_k$ is given by \eqref{beta} with $\tilde Y^i$ instead of $Y^i$. This equation is same with the vacuum equation of the pure 
ABJM theory without mass deformation except for the newly introduced coordinate $\xi$. Since $Y^i$ is periodic, so is $\tilde Y^i$. In the pure ABJM theory, we know that the only regular
periodic solutions satisfying the vacuum equation \eqref{tildeyeq} are
constant diagonal matrices, namely,
\begin{equation}
        \tilde Y^i_D = \textrm{diag}(y^i_1, y^i_2,\ldots,y^i_N).\label{diaY}
\end{equation}
Therefore, in the case with (\ref{sinqx}), the vacuum structure is essentially the same as that of the ABJM theory without mass deformation. The only difference is the exponential factor in \eqref{tildey}.
\subsection{$m(x) = m_0 + m_1 \sin{qx}$ 
case}\label{secm0m1}
As a next example, we add a constant mass  $m_0$ to \eqref{sinqx},
\begin{equation} \label{m0m1}
m(x) = m_0 + m_1 \sin{qx}
\end{equation}
with $m_0 > m_1 > 0$. This case may be considered as a perturbative modulation with a small parameter $m_1$ to the mABJM theory with a constant mass $m_0$. From the lesson in the previous example, it can be expected that the solution structure is not much different from that of the mABJM theory. In addition we consider $Y^a=0$ again to make the discussion simpler. Instead \eqref{xi}, we introduce another positive definite function $\xi(x)$ satisfying
\begin{equation} \label{eta}
\frac{d\xi}{dx} = e^{2m_0 \xi - 2\int m\,dx}.
\end{equation}
Then by integration, we obtain
\begin{align}\label{exi}
e^{-2m_0\xi} = - 2 m_0 c \int^x  \, e^{-2 \left(m_0 x' - \frac{m_1}{q} \cos qx'\right)}dx',
\end{align}
where we used the relation 
$e^{-2 \int^x dx' m(x')}= c\, e^{-2 \left(m_0 x - \frac{m_1}{q} \cos qx\right)}$
with an integration constant $c$.
Plugging \eqref{exi} into \eqref{eta} and taking the inverse, we obtain \begin{align}\label{dxdxi}
\frac{dx}{d\xi} &= 2m_0\int_x^\infty  e^{-2 m_0(x'-x) + \frac{2m_1}{q}\left(\cos qx' - \cos qx\right) }dx'.
\end{align}
Here we considered the integration range $(x,\infty)$ to guarantee $\xi = x$ up to a constant translation for $m_1=0$ case. 
From \eqref{dxdxi}, one can easily show that $\frac{d\xi}{dx}$
is a periodic function of $x$ with period $2\pi/q$ by changing the integration variable as $x'\to x' + \frac{2\pi}{q}$. Thus (\ref{eta}) implies that $\xi$ is nothing but $x$ modulated by a periodic function of $\mathcal{O}\left(m_1\right)$. Like (\ref{tildey}) in the previous example, we make use of the following redefinition of the scalar field: 
\begin{equation} \label{tildey2}
        Y^i = \left( \frac{d\xi}{dx} \right)^{1/2} \tilde Y^i
            = e^{m_0 \xi - \int m dx}\, \tilde Y^i.
\end{equation}
Then \eqref{veq13} becomes
\begin{equation} \label{tildeyeq2}
\frac{d}{d\xi} \tilde Y^i - \tilde\beta^{ki}_{k}+ m_0 \tilde Y^i = 0.
\end{equation}
This equation is the same as the vacuum
equation in the mABJM theory with a constant mass $m_0$ . Now the argument goes exactly the same
as that of the previous example. Since $Y^i$ is periodic, so is $\tilde Y^i$.
For the constant mass case, we know that the only regular
periodic solutions satisfying the vacuum equation \eqref{tildeyeq2} are
GRVV constant solutions \eqref{mABJMv}, i.e., $\frac{d}{d\xi} \tilde Y^i =0$. 
Therefore, we conclude that vacuum
solutions for this mass-function are given by GRVV constant matrices multiplied by the exponential 
factor in \eqref{tildey2}.

\subsection{General structure of periodic vacuum solutions}

Now we discuss the general vacuum solutions with nonvanishing $Y^a$ and $Y^i$ for generic periodic mass-functions, \begin{align}
m(x) = m_0 +   {\hat m}(x)~, 
\end{align} 
where  $m_0$ is a positive constant and $  {\hat m}(x)$ is a periodic function with a period $\tau$ satisfying $\int_0^\tau   {\hat m}(x) = 0$. Thus $m_0$ denotes the reference value of the mass-function $m(x)$.
In the previous subsection, we obtained the vacuum solutions for $Y^i$. For the vacuum solution of $Y^a$ satisfying the vacuum equation in \eqref{veq12}, we simply replace $m$ by $-m$ for $Y^i$ in \eqref{tildey2}.  In order to satisfy \eqref{veq14} for the nonvanishing $Y^a$ and $Y^i$ case, one needs to rearrange block-diagonal matrices as given in \eqref{mABJMv} for the constant mass deformation. Specific details are described below.

 Similarly to \eqref{tildey}, we  introduce  $\tilde{Y}^A$ as, 
\begin{align}\label{tildey3}
Y^a =  e^{-m_0(\eta-x) +\Lambda(x) }\, \tilde Y^a, \qquad
Y^i = e^{m_0(\xi-x) - \Lambda(x)}\, \tilde Y^i,
\end{align}
where $\Lambda(x) \equiv \int^x\hat m(x')dx'$ and  we define monotonically increasing  functions, $\eta$ and  $\xi$, satisfying the following differential equations:
\begin{align}\label{deta}
\frac{d\eta}{dx} =e^{-2m_0(\eta-x) +2\Lambda(x)},
\qquad
\frac{d\xi}{dx} = e^{2m_0(\xi-x) -2 \Lambda(x)}.
\end{align}
Inserting \eqref{tildey3} into the vacuum equations in \eqref{veq12} and \eqref{veq13}, we obtain the BPS equations for $\tilde Y^A$  as
\begin{align}\label{tilBPS}
\frac{d}{d\eta}\tilde Y^{a} - \tilde\beta^{ca}_{c}- m_{0} \tilde Y^a = 0,
\qquad
\frac{d}{d\xi}\tilde Y^{i}-  \tilde \beta^{ki}_{k}+ m_0 \tilde Y^i = 0,
\end{align}
where ${\tilde{\beta}}^{AB}_C$ is given by \eqref{beta} with $\tilde{Y}^A$ replacing $Y^A$. 
From the relations in \eqref{tildey3}, we notice that since $Y^A$'s are periodic, so are $\tilde Y^A$'s if  $\frac{d\eta}{dx}$ and $\frac{d\xi}{dx}$ are periodic. In order to show the periodicity  of  $\frac{d\eta}{dx}$ and $\frac{d\xi}{dx}$, we use the same method described in the previous subsection. From (\ref{deta}), one can find $e^{2 m_0 \eta}$ and $e^{2 m_0 \xi}$ by solving the differential equations. Plugging the expressions into (\ref{deta}) again, then we obtain 
\begin{align}\label{dxde}
\frac{dx}{d\eta} &= 2m_0\int_{-\infty}^x  e^{2 m_0(x'-x) -2\big(\Lambda(x) - \Lambda(x')\big)}dx',
\nn \\
\frac{dx}{d\xi} &= 2m_0\int_x^\infty  e^{-2 m_0(x'-x) + 2\big(\Lambda(x) - \Lambda(x')\big) }dx'.
\end{align}
When $\hat m = 0$, one expect that the BPS equations in \eqref{tilBPS} reduce to those replaced by  $ \tilde Y^A\to Y^A $ and $\eta, \xi \to x$. This is guaranteed by the integration range in \eqref{dxde}. Also one can easily check that $\frac{d\eta}{dx}$ and $\frac{d\xi}{dx}$ are periodic functions using $\Lambda(x) = \Lambda(x+\tau)$ and changing the integration variable from $x'$ to $x'' = x' + \tau$.

By using the functions $\eta$, $\xi$ and the field redefinition, we can find the general solutions for arbitrary periodic mass-functions. As we argued in the specific example in the previous section, the only regular periodic solutions satisfying the vacuum equations \eqref{tilBPS} are GRVV solutions \eqref{mABJMv}, which implies $\frac{d}{d\eta}\tilde Y^{a} =0$ and $\frac{d}{d\xi}\tilde Y^{i}=0$. Therefore, the explicit vacuum solutions for $\tilde Y^A$ in \eqref{tilBPS} with constraints \eqref{veq14} with $Y^A$ replaced by $\tilde Y^A$ are exactly same with those  in \eqref{mABJMv}.
From these constant vacuum solutions, $Y^a$ and $Y^i$ are read by attaching different exponential factors in \eqref{tildey3}.  We  write down the general solutions  for the nonvanishing $m_0$ case as follows:
\begin{align}\label{PsolGRVV}
&Y^a =\Sigma_+ (x)\,\tilde{Y}^a_{{m}_0} =  
\left(\frac{dx}{d\eta}\right)^{-1/2} \,\tilde{Y}^a_{{m}_0}  ~,\nonumber \\
&Y^i = \Sigma_-(x) \,\tilde{Y}^i_{m_0}  =  
\left(\frac{dx}{d\xi}\right)^{-1/2} \tilde{Y}^i_{m_0}  ~,
\end{align}    
where $\frac{dx}{d\eta}$ and $\frac{dx}{d\xi}$ are given in \eqref{dxde}. Here $\bar{Y}^a_{m_0}$ and $\bar{Y}^i_{m_0}$ are the GRVV matrices, whose explicit form is given in (\ref{mABJMv}) with replacing $m$ with $m_0$. Also, $\beta^{ja}_{i} =  \beta^{bi}_{a} =\beta^{ab}_{i}=\beta^{ij}_{a}=0$ automatically.

Especially for $m_0=0$ case, one has to start from \eqref{tildey3}  and \eqref{deta}, i.e., 
\begin{align}\label{tildey4}
Y^a =  e^{ \Lambda(x)}\, \tilde Y^a, \qquad
Y^i = e^{- \Lambda(x)}\, \tilde Y^i.
\end{align} 
Then we obtain 
\begin{align}\label{tilBPS2}
\frac{d}{d\eta}\tilde Y^{a} - \tilde\beta^{ca}_{c} = 0,
\qquad
\frac{d}{d\xi}\tilde Y^{i}-  \tilde \beta^{ki}_{k}=0,
\end{align}
where $\eta$ and $\xi$ are new coordinates given by 
\begin{align}
\eta(x) = \int^x e^{2\Lambda(x')}dx',\
\qquad \xi(x) = \int^x e^{-2\Lambda(x')}dx'.
\end{align}
As we discussed in the subsection \ref{mx=m1s}, $\tilde Y^a$ and $\tilde Y^i$ are periodic functions and then the  regular solutions should satisfy $\tilde\beta^{ca}_{c}=0$ and $\tilde \beta^{ki}_{k}=0$.
Therefore, we can find the general solutions as \begin{align}\label{PsolDiag}
&Y^a =   e^{ \Lambda(x')} \tilde{Y}^a_{D}~~,~~Y^i =   e^{-\Lambda(x')}\tilde{Y}^i_{D}~,
\end{align}
where $\tilde Y_D^a$ and $\tilde Y^i_D$ are diagonal matrices given in \eqref{diaY}.

\subsection{Classical limit of vevs  for general periodic vacuum solutions}
The general periodic vacuum solutions (\ref{PsolDiag}) and (\ref{tildey3}) are induced by periodic mass deformations, which correspond to a spatial coordinate dependent source to the pure ABJM theory and mABJM theory, respectively. These sources cause nonvanishing expectation values of gauge invariant operators. It would be interesting to evaluate those vevs for the spatially modulated mass deformations. Holographic approach for the calculations of the vevs was investigated in \cite{Gauntlett:2018vhk,Arav:2018njv}. 
We limit our consideration to  classical vacua only in the ${\cal N} = 3$ ImABJM model  in this work.

First, we discuss the $m_0=0$ case with the corresponding solution (\ref{PsolDiag}). Plugging the vacuum solution into (\ref{Operators}), the classical vevs of gauge invariant operators are obtained as follows,
\begin{align}\label{OpDiagonal}
&\langle\mathcal{O}^{\Delta=1}\rangle_0  = e^{2 \Lambda(x)} \text{tr}\left( \tilde{Y}_D^a \tilde{Y}_{aD\,}^\dagger \right)-e^{-2 \Lambda(x)} \text{tr}\left( \tilde{Y}^i_D \tilde{Y}_{iD\,}^\dagger \right),~
\nonumber\\
&\langle\mathcal{O}^{\Delta=2}\rangle_0 =0,
\nonumber\\
&\langle\mathcal{O}^K \rangle_0 =   e^{2 \Lambda(x)} \text{tr}\left( \tilde{Y}_D^a \tilde{Y}_{aD\,}^\dagger \right)+  e^{-2 \Lambda(x)} \text{tr}\left( \tilde{Y}_D^i \tilde{Y}_{iD}^\dagger \right),
\end{align}
where $\langle\mathcal{O}\rangle_0$ denotes the classical vev for an operator ${\cal O}$. Here we notice that the mass deformation with vanishing $m_0$ cannot generate 
$\langle\mathcal{O}^{\Delta=2}\rangle_0 $, since $\beta^A_{BC}|_{Y= Y_0}$'s with the vacuum solution (\ref{PsolDiag}) are always vanishing.

In the next, we discuss the $m_0\neq 0$ case. This case gives us more nontrivial classical vevs for the gauge invariant operators. 
Plugging the solution (\ref{tildey3}) into the operator expressions (\ref{Operators}), the classical vevs become 
\begin{align}
&\langle\mathcal{O}^{\Delta=1}\rangle_0    = (\Sigma_+)^2\, \text{tr}\left( \tilde{Y}_{m_0}^a \tilde{Y}_{m_0\,a}^\dagger \right)-(\Sigma_-)^2 \text{tr}\left( \tilde{Y}^i_{m_0} \tilde{Y}_{m_0\,i}^\dagger \right),
\nn \\
&\langle\mathcal{O}^{\Delta=2}\rangle_0  = -4 m_0\left[   (\Sigma_+)^4  \text{tr}\left( \tilde{Y}_{m_0}^a \tilde{Y}_{m_0\,a}^\dagger \right)+ (\Sigma_-)^4   \text{tr}\left( \tilde{Y}^i_{m_0} \tilde{Y}_{m_0\,i}^\dagger \right)\right], \label{vev12K}
\\
&\langle\mathcal{O}^K \rangle_0  =  (\Sigma_+)^2 \text{tr}\left( \tilde{Y}_{m_0}^a \tilde{Y}_{m_0\,a}^\dagger \right)+ (\Sigma_-)^2 \text{tr}\left( \tilde{Y}^i_{m_0} \tilde{Y}_{m_0\,i}^\dagger \right),\nn
\end{align}     
where $\Sigma_\pm(x)$ is defined in (\ref{PsolGRVV}). In addition, the traces of the scalar vacuum fields are given by
\begin{align}
&\text{tr}\left( \tilde{Y}_{m_0}^a \tilde{Y}_{m_0\,a}^\dagger \right) = \frac{m_0 k}{2\pi}\sum_{n=0}^{\infty}
n(n+1)N_n,\\
&\text{tr}\left( \tilde{Y}^i_{m_0} \tilde{Y}_{m_0\,i}^\dagger \right)=\frac{m_0 k}{2\pi}\sum_{n=0}^{\infty}
n(n+1)N_n'.
\end{align}
For the classical vevs in \eqref{vev12K}, we are considering a finite $N$ in the field theory side. In order to compare these vevs with those in the corresponding gravity theory, which are expected to match the vevs in the strong coupling limit, one has to take into account the serious quantum corrections. However, for some chiral primary operators, for instance ${\cal O}^{\Delta=1,2}$ in \eqref{Operators}, the classical vevs in the mABJM theory match the holographic vevs from the gravity theory through the large 
$N$ limit~\cite{Jang:2019pve}. This is a quite remarkable properties of vevs of the CPOs between weak and strong coupling limits of the mABJM theory.

Now let us consider classical vevs of regular configuration in the representative examples. As a first example, we consider $m(x)= m_1 \sin q x$ analyzed in the subsection \ref{mx=m1s}, where $m_1>0$ and $q= 2\pi n /(x_R - x_L)$ for some integer $n$. 
The solution  is given by (\ref{PsolDiag}). The explicit form of the solution is  written as 
\begin{align}\label{sol_Sinqx}
Y^a =c_a e^{\frac{-m_1}{q}\cos qx  } \tilde{Y}_D^a,
\qquad
Y^i =c_i e^{\frac{-m_1}{q}\cos qx} \tilde{Y}_D^i,
\end{align}
where $c_a$ and $c_i$ are arbitrary constants.
 When we take very small value of $m_1(\ll q)$ with $ \sum_{a=1}^2 \left( c_a\,\text{tr} \tilde{Y}_D^a \tilde{Y}_{D\,a}^\dagger \right)=\sum_{i=3}^4 \left( c_i\,\text{tr} \tilde{Y}_D^i 
\tilde{Y}_{D\,i}^\dagger \right)= -\tilde c $, the classical vevs become
\begin{align}\label{sinDiag}
\langle\mathcal{O}^{\Delta=1}\rangle_0  \sim  \tilde{c} \,m_1 \cos q x~~,~~\langle\mathcal{O}^{\Delta=2}\rangle_0 = 0~~,~~\langle\mathcal{O}^{K}\rangle_0 \sim -2\tilde c.
\end{align}
Even though we are working with classical configurations, this limit is similar to the gravity dual of a  spatially modulated deformation of the ABJM theory, except for the dimension 2 operator in \cite{Gauntlett:2018vhk,Arav:2018njv}. On the other hand, such a mismatch is quite natural because one need to take quantum corrections with the large $N$ limit to compare the operator expectation values in the both sides. So more complete analysis on the comparison should be considered. We leave the study as a future work~\cite{KKKK}.

Also, there is another type of deformation which has a nonvanishing reference value of the mass-function. One may take the mass-function as $m(x)=m_0 + m_1 \sin q x$, which was discussed in the subsection \ref{secm0m1}. For nonvanishing $m_0$, the corresponding solutions are given by (\ref{PsolGRVV}). 
We consider the spatial modulation as a perturbation around the GRVV vacua of the constant mass deformation. To do that,  we take a limit as $|{m_1}/{m_0}|\ll 1$. Then we obtain up to leading order of $m_1$ as 
\begin{align}
\Sigma_+(x) &=  
\left(\frac{dx}{d\eta}\right)^{-1/2}\sim 1- \frac{q \cos q x - 2 m_0 \sin q x}{4 m_0^2 + q^2}\, m_1, 
\nn \\
\Sigma_-(x) & =  
\left(\frac{dx}{d\xi}\right)^{-1/2}  \sim 1+ \frac{q \cos q x + 2 m_0 \sin q x}{4 m_0^2 + q^2}\, m_1.
\end{align}
Now we choose a vacuum which satisfies $\text{tr}\left( \tilde{Y}_{m_0}^a \tilde{Y}_{m_0\,a}^\dagger \right)=\text{tr}\left( \tilde{Y}^i_{m_0} \tilde{Y}_{m_0\,i}^\dagger \right)=m_0 \tilde{c}^2$, then the classical vevs become
\begin{align}
&\langle\mathcal{O}^{\Delta=1}\rangle_0    \sim -4m_0\tilde{c}^2  \frac{q \cos qx}{4 m_0^2 + q^2} m_1, 
\nn \\
&\langle\mathcal{O}^{\Delta=2}\rangle_0    \sim - 8 m_0^2 \tilde{c}^2 \left( 1 + \frac{8 m_0 \sin qx}{4 m_0^2 + q^2}\, m_1\right),\\
&\langle\mathcal{O}^{K}\rangle_0    \sim 2 m_0 \tilde{c}^2  \left( 1 +  \frac{4 m_0 \sin qx}{4 m_0^2 + q^2}\, m_1  \right). \nn
\end{align}
Unlike the diagonal case, the dimension 2 operator has nontrivial value. We also obtain various vacuum configurations induced by spatially modulated mass deformations, including singular mass configurations in Appendix \ref{singmass}.

\section{Conclusion}
A dual gravity solution of the ${\cal N} = 3$ ImABJM model with a spatially modulated mass-functions is conjectured as the SUSY Q-lattice solution in 11-dimensional supergravity. Using the holographic method, some features of strongly coupled limit of the ImABJM model were analyzed in \cite{Gauntlett:2018vhk,Arav:2018njv}. On the other hand, weakly coupled limit is another interesting research area. In this paper, we have investigated various aspects of the ImABJM models, which are useful in analyzing the weakly coupled limit of the models. Our result is composed of the construction of the ${\cal N} =1,2$ supersymmetric ImABJM models and general vacuum solutions for periodic mass-functions in the ${\cal N} = 3$ model. 

We found a general supersymmetric condition for  mass matrices with one spatial coordinate dependence. As special choices of the mass matrices and the supersymmetric parameters, we obtained the ${\cal N} =1$ and ${\cal N} = 2$  ImABJM models, 
where one needs three and two arbitrary mass-functions, respectively.  
In addition, for the ${\cal N} =3$ model, we discussed the supersymmetric vacuum energy conditions and constructed the general vacuum solutions for periodic mass-functions. We  showed that there are two types of vacuum solutions.  
For $\int_0^\tau dx\, m(x)=0$ case,  vacuum solutions for scalar fields become diagonal, while for $\int_0^\tau dx\, m(x)\ne 0$ case, those are proportional to the GRVV matrices. We constructed vevs of CPOs with conformal dimensions $\Delta = 1,2$ and discussed corresponding gravity duals.
As examples, we showed various vacuum solutions in Appendix B.

Since the ImABJM models allow arbitrary mass-functions, one can also try to obtain vacuum solutions for non-periodic mass-functions. For these cases, it seems that the energies of the vacuum solutions satisfying \eqref{vcon1} and \eqref{vcon2} are non-vanishing.  
This property may be understood by analyzing the supersymmetry algebra of the ImABJM models~\cite{KKKK}.  
As another future direction, it would be interesting to study gravity duals of the ${\cal N} =1,2$ ImABJM models,  since those include 3 and 2 arbitrary mass-functions, respectively. We expect such an arbitrariness is helpful in implementing more realistic applications.

\section*{\bf Acknowledgement}

This work was supported by the National Research Foundation of Korea(NRF) grant with grant number NRF-2019R1F1A1059220 (C.K. and O.K.), NRF-2019R1A2C1007396 (K.K.),  NRF-2019R1F1A1056815 (Y.K.) and NRF2017R1D1A1A09000951, NRF-2019R1A6A1A10073079 (O.K.). K.K acknowledges the hospitality at APCTP where part of this work was
done.

\vspace{0cm}

\appendix

\section{Derivation of \eqref{absrel}}\label{urels}
We consider the quantity in the supersymmetric variation of the fermion field 
in \eqref{fertr}. Except for the supersymmetric parameter $\omega_{AB}$ in \eqref{fertr}, the absolute square of the bosonic part is written as 
\begin{align}
|\delta^{[B}_A Y^{C]'} +s^{BC} G^{BC}_A|^2 =|\delta^{[B}_A Y^{C]'}|^2 + |G^{BC}_A|^2 + s^{BC} G^{BC}_A  (\delta^{[B}_A Y^{C]'})^\dagger +  ({\rm c.c.}).
\end{align}
Using the relations,
\begin{align}
&|\delta^{[B}_A Y^{C]'}|^2 = \delta^{[B}_A Y^{C]'} \delta_{[B}^A Y_{C]}^{\dagger '} = \frac{3}{2} |Y^{A'}|^2,  
\nn \\
&|G^{BC}_A|^2 =\frac{3}{2}\left(\hat V_{{\rm bos}} -\hat V_J\right),  
\end{align}
we obtain 
\begin{align}\label{YVV}
|Y^{A'}|^2 + V_{{\rm bos}} - V_J = \frac{2}{3} \left[|\delta^{[B}_A Y^{C]'} +s^{BC} G^{BC}_A|^2  -s^{BC} G^{BC}_A  (\delta^{[B}_A Y^{C]'})^\dagger +  ({\rm c.c.})\right].
\end{align}
We also get the relation 
\begin{align}\label{sGY}
s^{BC} G^{BC}_A  (\delta^{[B}_A Y^{C]'})^\dagger
&= -s\left[\left( G^{ac}_a   -  G_i^{ic} \right)Y_c^{\dagger'}- \left(  G_a^{ak}    -  G_i^{ik} \right)Y_k^{\dagger'} \right],
\end{align}
where we set the quantity $s_{AB} = s^{AB}$ to encode the projection \eqref{proj} as 
\begin{align}
s^{bc}= s^{jk} \equiv -1\equiv -s,\qquad 
s^{bi} = s^{ja} \equiv +1\equiv +s.
\end{align}
Here we introduced the letter $s$ for calculational convenience, which will be set as $s=1$ later. 
Using the definition \eqref{GABC},
we rewrite \eqref{sGY} as  
\begin{align}\label{sGY2}
s^{BC} G^{BC}_A  (\delta^{[B}_A Y^{C]'})^\dagger  = -\frac{3s}2\left[ \left(\beta^{ac}_a -\frac13 \beta_i^{ic} +  m Y^c\right)Y_c^{\dagger '} - \left(\frac13 \beta^{ak}_a - \beta_i^{ik} + m Y^k\right)Y_k^{\dagger '} \right].
\end{align} 
From \eqref{YVV},\eqref{sGY2}, $V_J = m' M_A^{~B},  Y_B^\dagger Y^A$, and setting  $s=1$, we finally obtain 
\begin{align}
|Y^{A'}|^2 + V_{{\rm bos}}&=\frac{2}{3}|\delta^{[B}_A Y^{C]'} +s^{BC} G^{BC}_A|^2 
\nn \\
&~~+\Big(\frac12\beta^{ac}_a Y_c^{\dagger}  + \frac12\beta^{ik}_i Y_k^{\dagger} -\frac13  \beta^{ic}_i Y_c^\dagger
+ m M_A^{~B} Y_B^\dagger Y^A\Big)'.
\end{align}

\section{Various Vacuum Configurations with Periodic Mass-functions }\label{singmass}

This section provides various examples of the general vacuum solution obtained in section \ref{pvacsol}. We take into account singular mass-functions as well as regular ones. From the previous analysis, we found that there exist two classes of solutions, which are given by (\ref{PsolGRVV}) and (\ref{PsolDiag}). In order to reflect such classification, we use following ansatz to describe explicit examples,
\begin{align}
&Y^a = P(x) \tilde{Y}_D^a ~~,~~Y^a = Q(x) \tilde{Y}_D^i ~~~~~: \text{Diagonal Type,}\\
&Y^a = P(x) \tilde{Y}^a_{\tilde m=1}~~,~~ Y^i = Q(x) \tilde{Y}^i_{\tilde m=1} ~~: \text{GRVV Type}~,
\end{align}
where we took the scaled GRVV matrices $\tilde{Y}^a_{\tilde m=1}$ and $\tilde{Y}^i_{\tilde m=1}$ for convenience. They are given by (\ref{mABJMv}) with taking $\tilde m\equiv \sqrt{\frac{mk}{2\pi}}=1$, so $P(x)$ and $Q(x)$ become dimensionful functions. Together with this ansatz, the BPS equations (\ref{veq12}) and (\ref{veq13}) can be written in the following form,
\begin{align}
&P'(x) -m(x) P(x) = 0,~~~~~~~~~~~~Q'(x) + m(x) Q(x) = 0 ~~~~~~~~~ :\text{Diagonal Type,}\label{D_BPSeq}\\
&P'(x) + P(x)^3 -m(x) P(x) = 0,~Q'(x)- Q(x)^3 + m(x) Q(x) = 0  ~:\text{GRVV Type.}\label{GR_BPSeq}
\end{align}

\begin{figure}[ht!]
\centering
    \subfigure[ ]
    {\includegraphics[width=7.8cm]{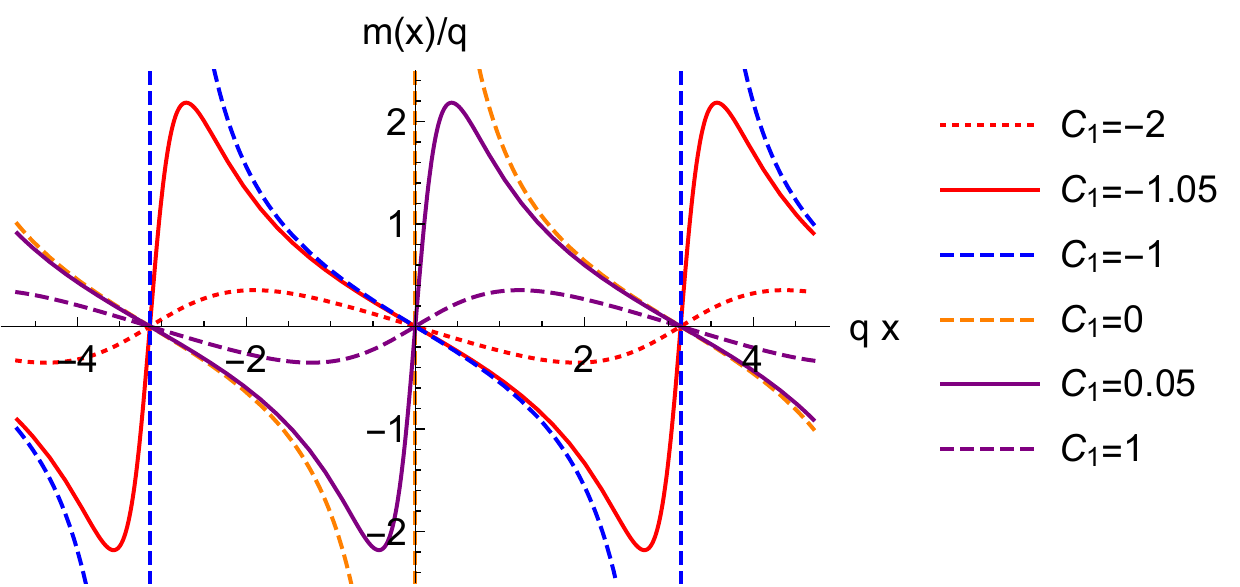}  }
    \hskip0cm
    \subfigure[ ]
    {\includegraphics[width=7.8cm]{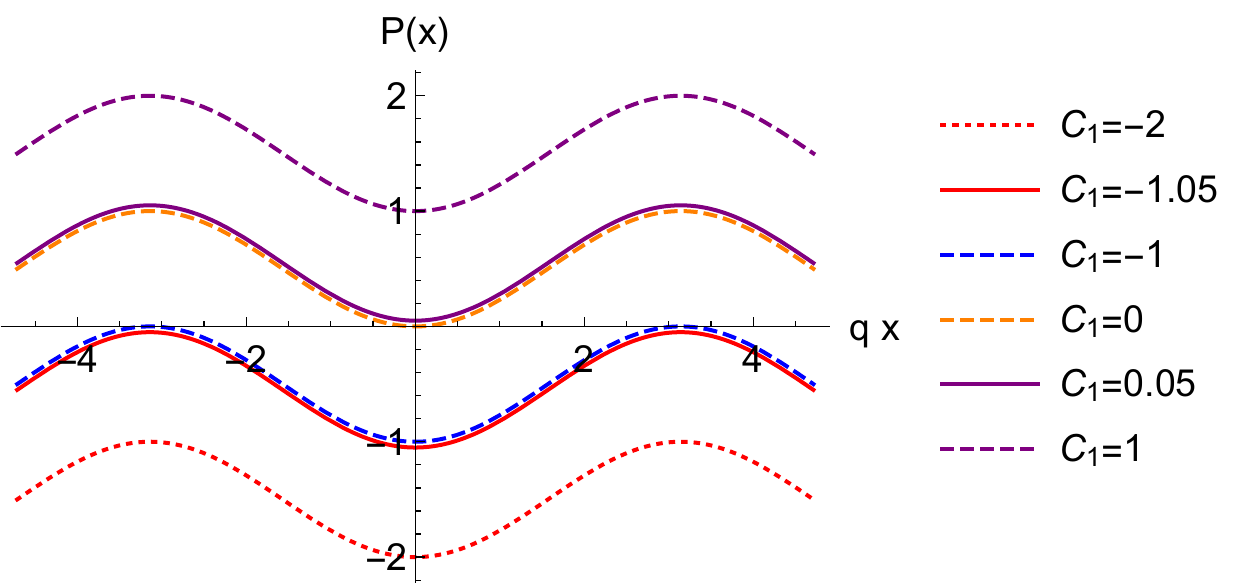}  }
    \hskip0.2cm

    \caption{Diagonal Types with $m(x)=\frac{q \sin (q x)}{2 \left(C_1+\sin ^2\left(\frac{q x}{2}\right)\right)}$. }
      \label{fig:beta01}
\end{figure}

As we expected from section \ref{pvacsol}, the diagonal type equation (\ref{D_BPSeq}) implies $\int_0^\tau m(x') dx'=0$ for periodic $P(x)$ and $Q(x)$. Also one can notice that $P(x)Q(x)$ must be constant by plugging $m(x)$ into the other equation. In addition, (\ref{D_BPSeq}) tells how the regularity of mass-function is related to the scalar configurations. If $P(x)$ or $1/Q(x)$ vanishes, the corresponding mass-function must be a singular function. Therefore, some regular scalar field configurations can exist even for singular mass configurations. As a first example, if we choose a mass-function with $m(x)=\frac{q \sin (q x)}{2 \left(C_1+\sin ^2\left(\frac{q x}{2}\right)\right)}$, then the corresponding $P(x)$ and $Q(x)$ are given as the following three cases:
\begin{align}
&\text{Case 1}: P(x)= \left(C_1+\sin ^2 \frac{q x}{2 }\right),~~Q(x)=0, \\
&\text{Case 2}: P(x)= 0,~~ Q(x) =  \left(C_1+\sin ^2 \frac{q x}{2 }\right)^{-1},  \\
&\text{Case 3}:P(x) = 1/Q(x) =  \left(C_1+\sin ^2 \frac{q x}{2 }\right),
\end{align} 
where $C_1$ is a constant which controls singular behavior of the scalar fields and the mass-function. When $C_1$ is very large, the mass-function becomes $m(x)\sim \frac{q}{2C_1} \sin q x$, this limit is comparable to the case of (\ref{sol_Sinqx}).  The classical vevs of the dimension 1 operators are given as follows:
\begin{align}
&\text{Case 1}:~~ \mathcal{O}^{\Delta=1}=\left(C_1+\sin ^2 \frac{q x}{2 }\right)^2\, \text{tr}\left( \tilde{Y}_D^a \tilde{Y}_{D\,a}^\dagger \right),  \\
&\text{Case 2}:~~ \mathcal{O}^{\Delta=1}=-\left(C_1+\sin ^2 \frac{q x}{2 }\right)^{-2}\, \text{tr}\left( \tilde{Y}_D^i \tilde{Y}_{D\,i}^\dagger \right), \\
&\text{Case 3}:~~ \mathcal{O}^{\Delta=1}=\left(C_1+\sin ^2 \frac{q x}{2 }\right)^2\, \text{tr}\left( \tilde{Y}_D^a \tilde{Y}_{D\,a}^\dagger \right)-\left(C_1+\sin ^2 \frac{q x}{2 }\right)^{-2}\, \text{tr}\left( \tilde{Y}_D^i \tilde{Y}_{D\,i}^\dagger \right),
\end{align}
while $\mathcal{O}^{\Delta=2}$ vanishes identically. In general, the mass-function or the scalar fields becomes singular, when $-1<C_1<0$. Since the case 2 and 3 do not have regular field configurations, the only allowed case is the case 1 corresponding to vanishing $Q(x)$. We plot the mass-function and $P(x)$ for this case in Figure \ref{fig:beta01}.

\begin{figure}[ht!] 
\centering
    \subfigure[ ]
    {\includegraphics[width=7.8cm]{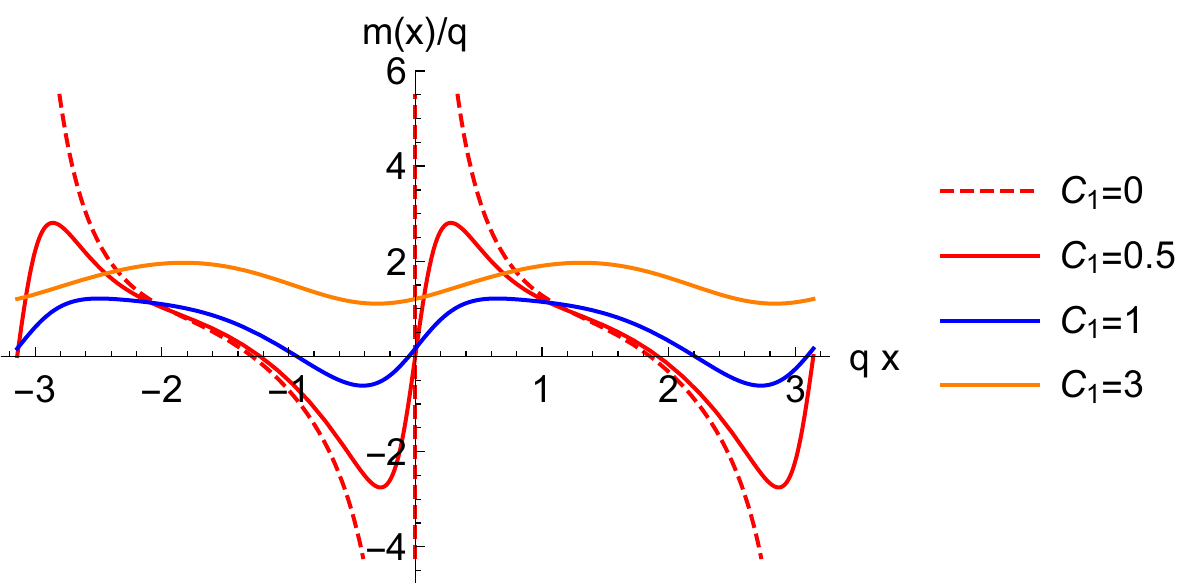}  }
    \hskip0cm
    \subfigure[ ]
    {\includegraphics[width=7.8cm]{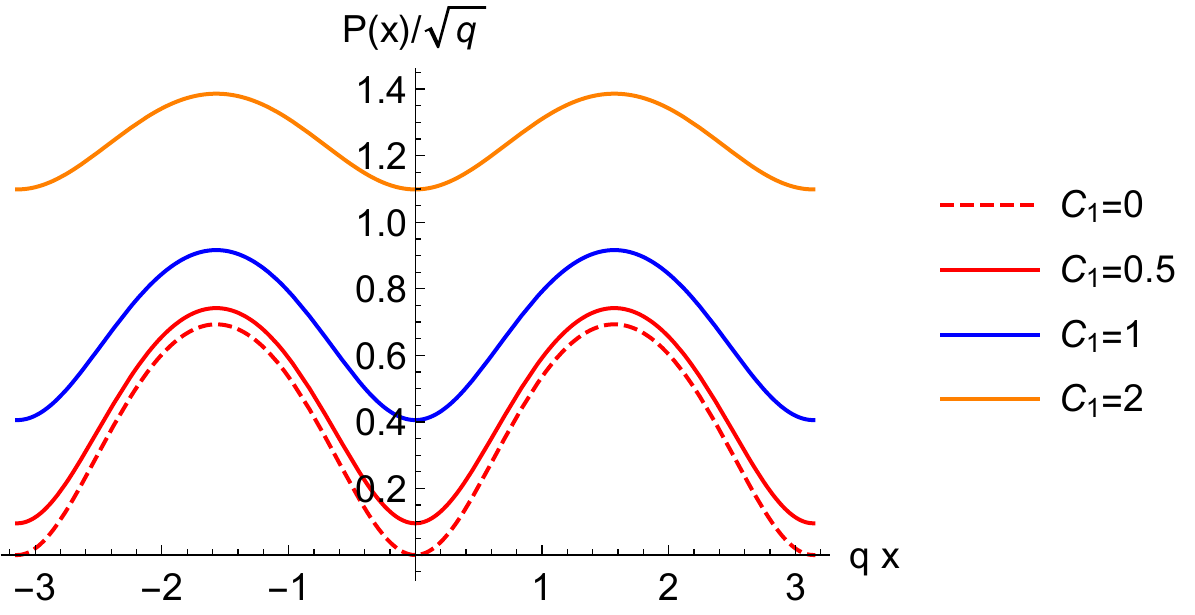}  }
    \hskip0.2cm

    \caption{GRVV Types with mass-function $m(x)=q \frac{\left(\left(\sin ^2(q x)-C_1\right){}^3+\sin (2 q x)\right)}{\sin ^2(q x)-C_1}$.  }\label{fig:GRVV01}
\end{figure}

Now let us consider examples of the GRVV type. The corresponding BPS equations are given by (\ref{GR_BPSeq}). First we consider a mass-function which is given by 
\begin{align}\label{mass1}
m(x)=q \frac{\left(\left(\sin ^2(q x)-C_1\right){}^3+\sin (2 q x)\right)}{\sin ^2(q x)-C_1}.
\end{align}
Then the $P(x)$ has the following form:
\begin{align}
P(x) =\sqrt{q} \left(\sin ^2(q x)-C_1\right).
\end{align}
One interesting configuration is the above solution with $Q(x)=0$. Then the scalar field configuration are always regular even for a singular mass-function. We plot $P(x)$ and the mass-function (\ref{mass1}) in Figure. \ref{fig:GRVV01}.

\begin{figure}[ht!] 
\centering
    \subfigure[ ]
    {\includegraphics[width=7.8cm]{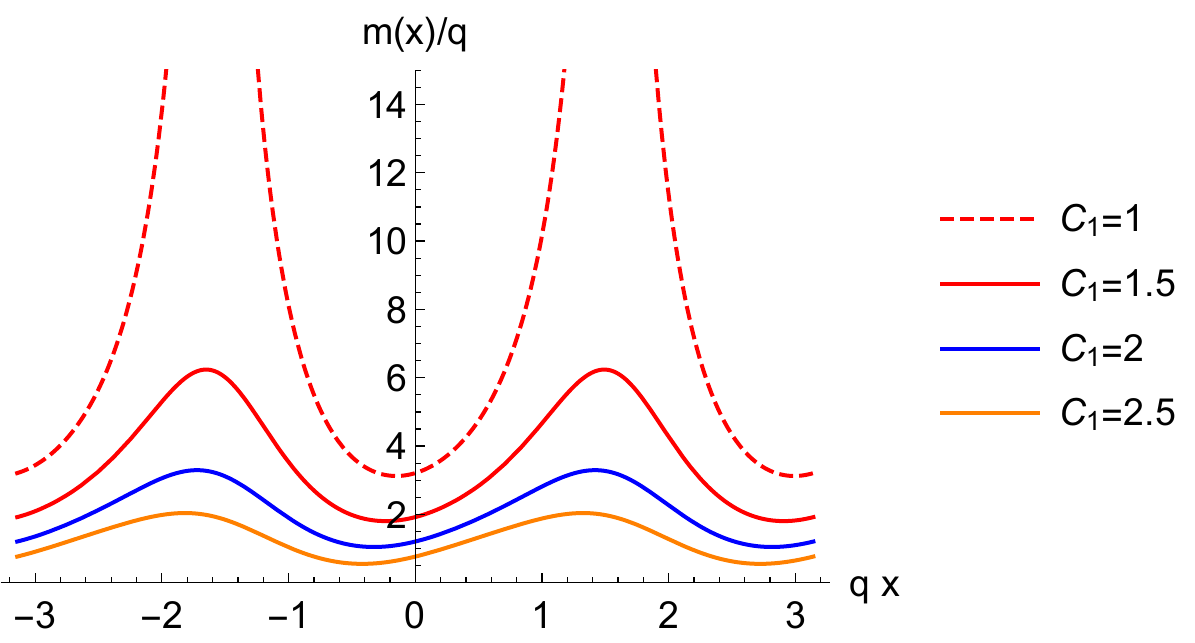}  }
    \hskip0cm
    \subfigure[ ]
    {\includegraphics[width=7.8cm]{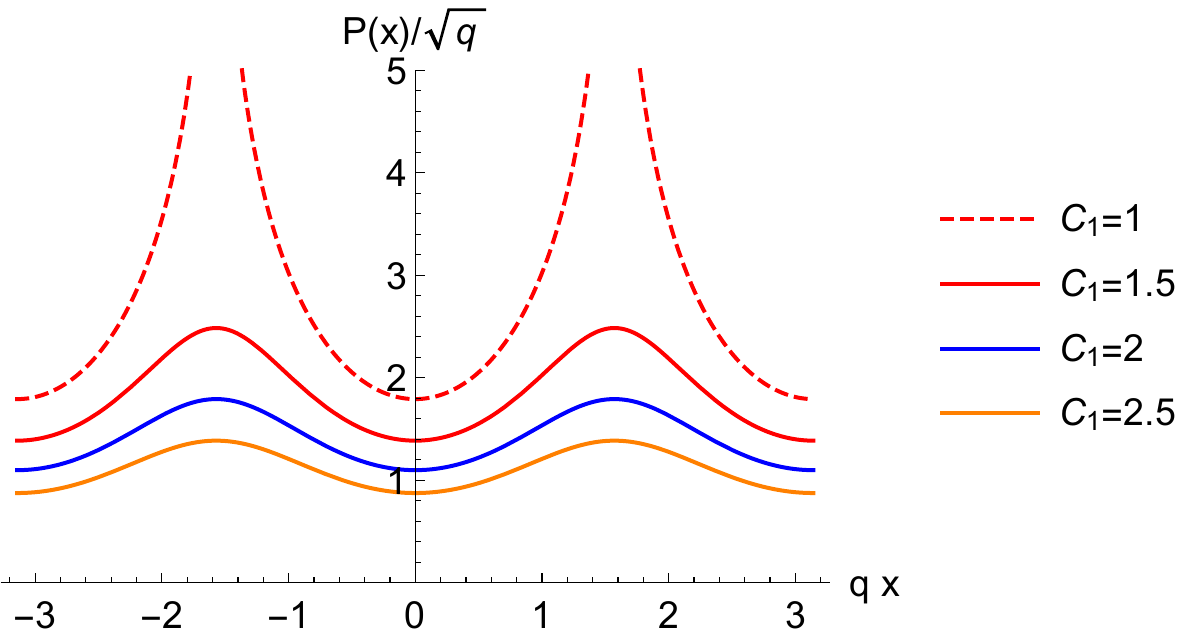}  }
    \hskip0.2cm

    \caption{GRVV Types with mass-function $m(x)=\frac{q \left(\log ^3\left(\frac{1}{6} \left(C_1-\sin ^2q x\right)\right)-\frac{2 \sin (2 q x)}{2 C_1+\cos (2 q x)-1}\right)}{\log \left(\frac{1}{6} \left(C_1-\sin ^2(q x)\right)\right)}$.
  }\label{fig:GRVV02}
\end{figure}

As a second example of GRVV type solutions, we consider a mass-function, which is positive in the whole region. For such a case, we take the following mass-function with $Y^i=0$, i.e, $Q(x)=0$:
\begin{align}
m(x)=\frac{q \left(\log ^3\left(\frac{1}{6} \left(C_1-\sin ^2q x\right)\right)-\frac{2 \sin (2 q x)}{2 C_1+\cos (2 q x)-1}\right)}{\log \left(\frac{1}{6} \left(C_1-\sin ^2(q x)\right)\right)}.~~
\end{align} 
Then the scalar field has the following form: 
\begin{align}
P(x) = -\sqrt{q} \log \left(\frac{1}{6} \left(C_1-\sin ^2(q x)\right)\right),
\end{align}
where $C_1$ is again control parameter of the regularity. For $C_1>1$, the vacuum solution and the mass-function are regular functions. When $C_1=1$, the mass-function and the scalar field diverge at $x= \frac{ \pi}{q}\left( n+ \frac{1}{2}\right)$, where $n$ is an integer. These configurations looks like periodic potentials. We plot various mass-functions and solutions in Figure \ref{fig:GRVV02}.

\begin{figure}[ht!] 
\centering
    \subfigure[ ]
    {\includegraphics[width=4.8cm]{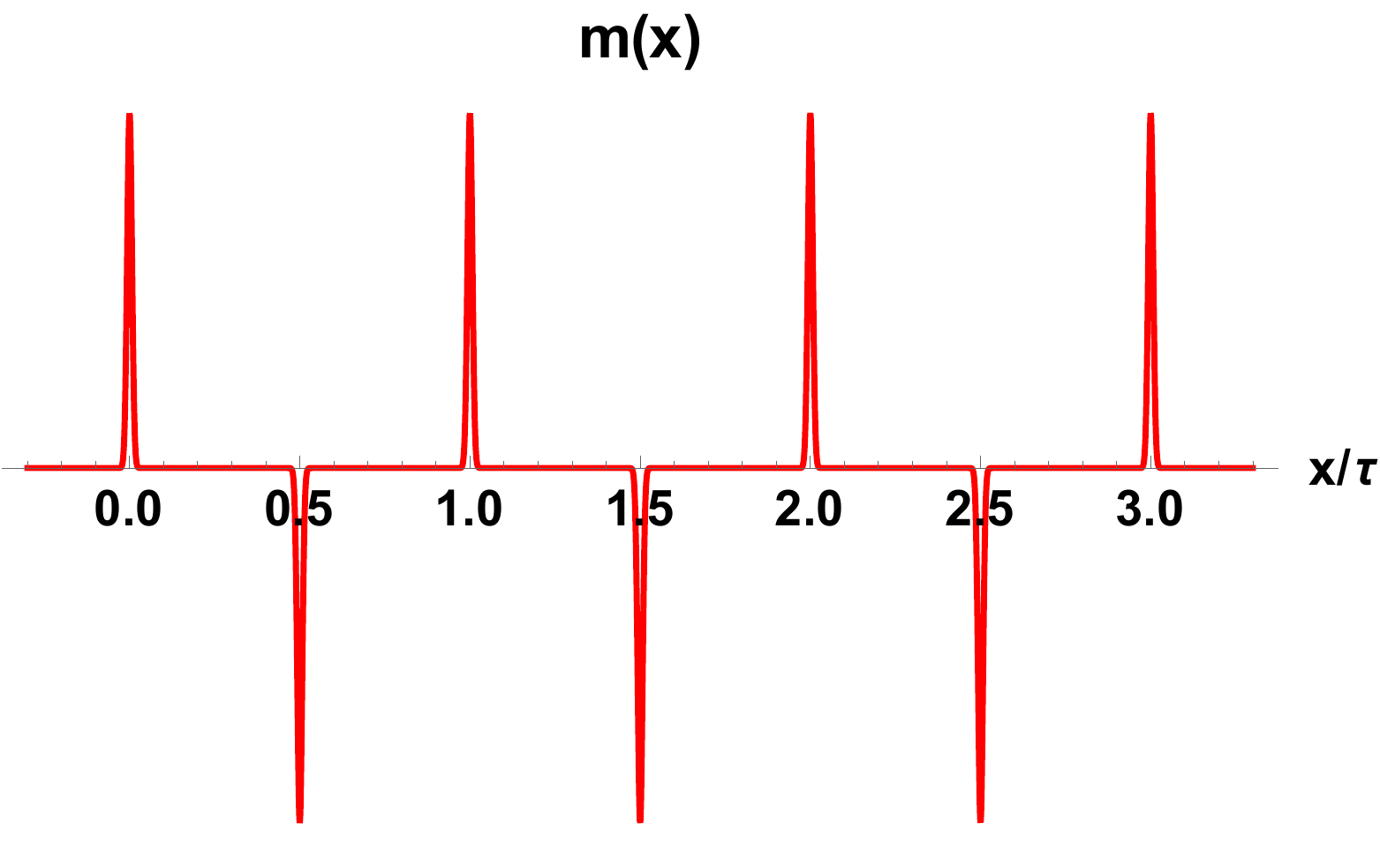}  }
    \hskip0cm
    \subfigure[ ]
    {\includegraphics[width=4.8cm]{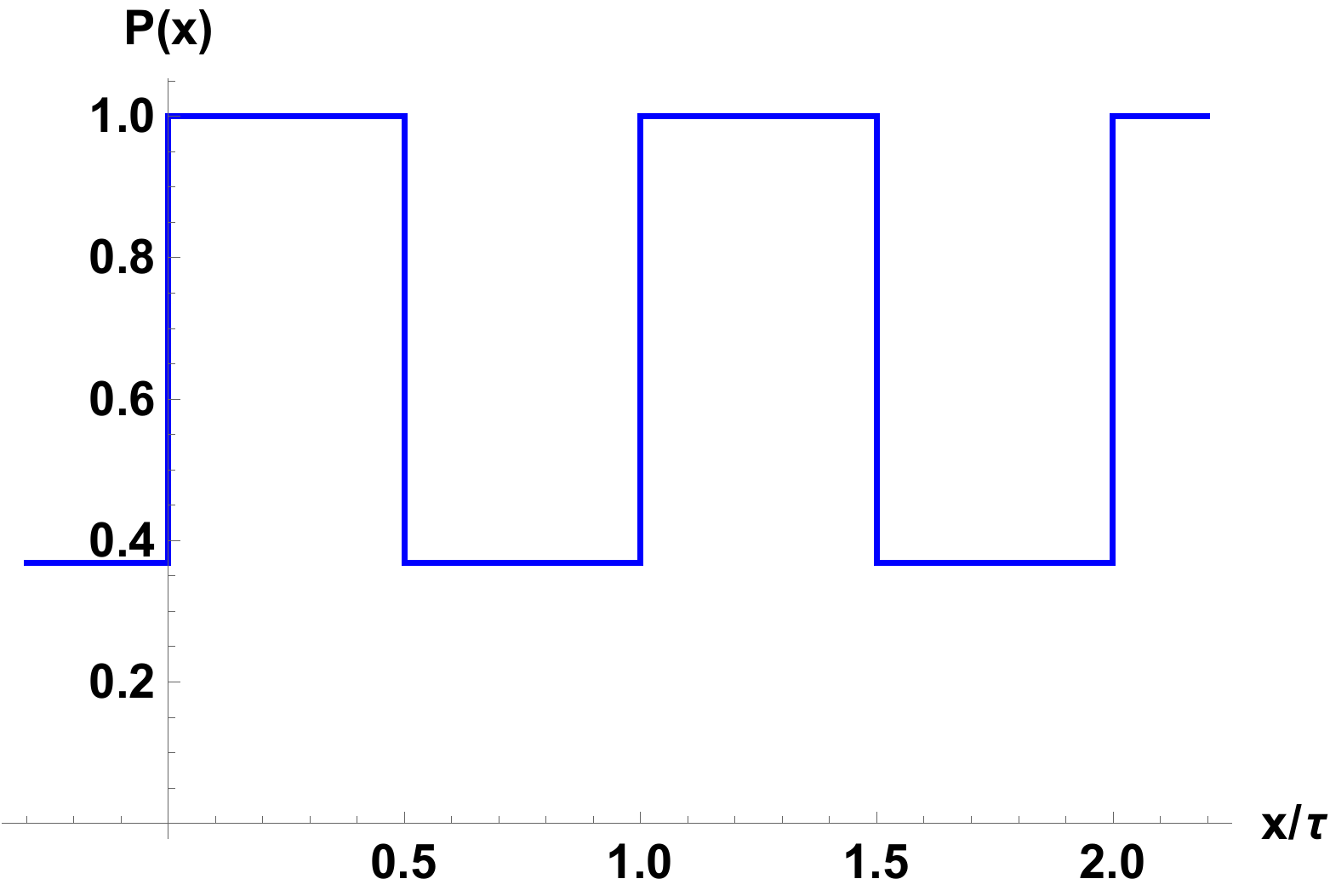}  }
    \hskip0cm
    \subfigure[ ]
    {\includegraphics[width=4.8cm]{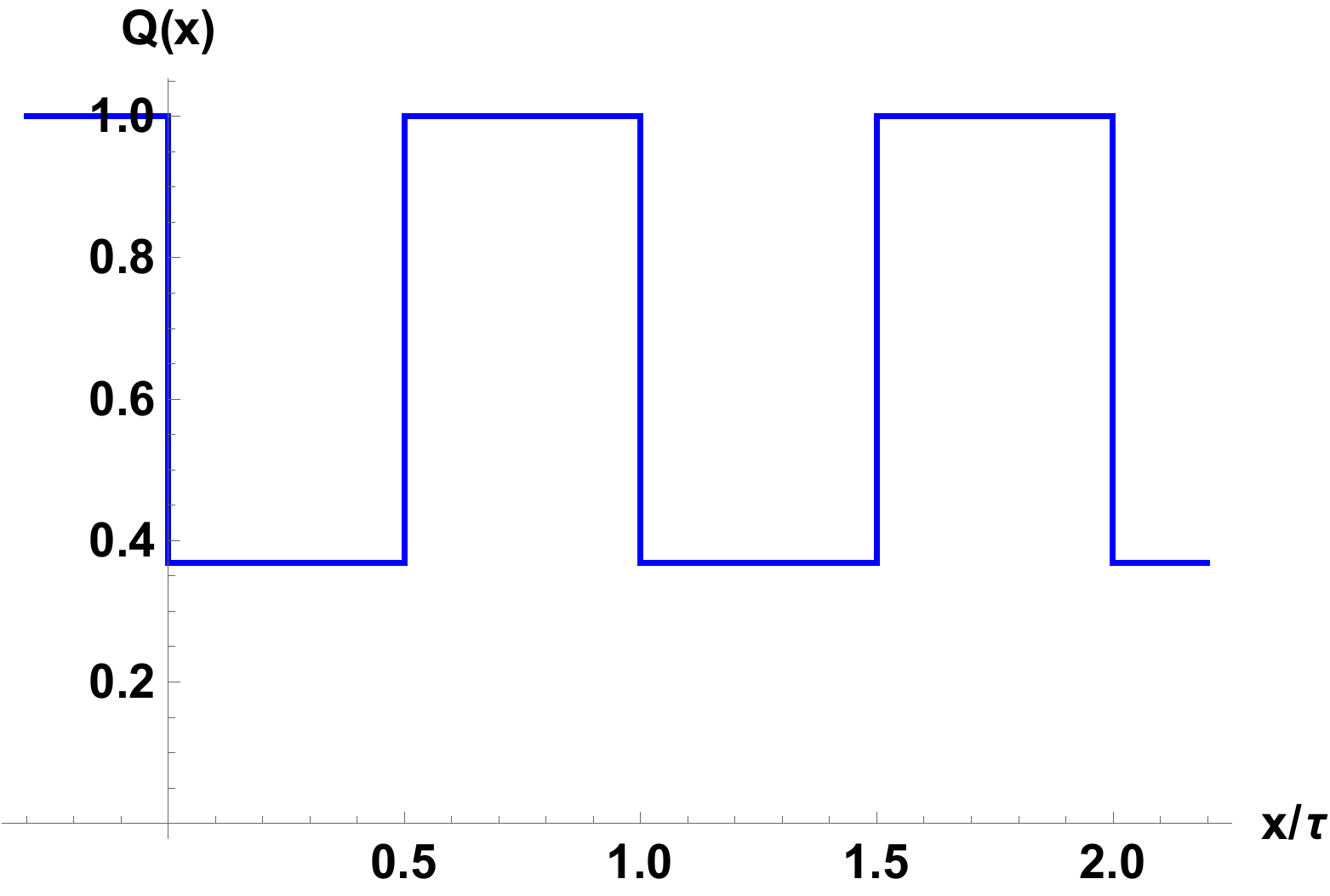}  }
    \hskip0cm\\
    \subfigure[ ]
    {\includegraphics[width=4.8cm]{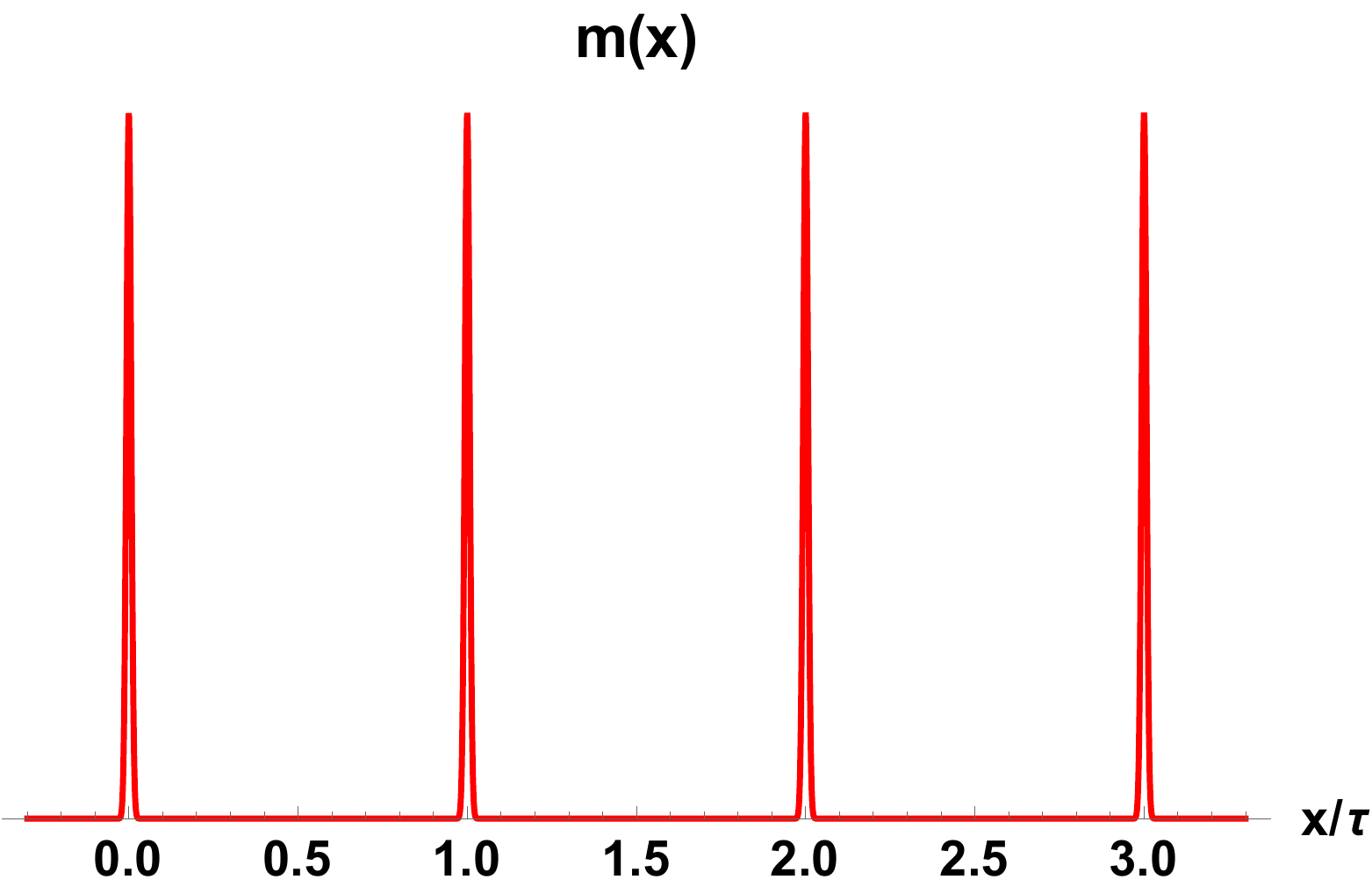}  }
    \hskip0cm
    \subfigure[ ]
    {\includegraphics[width=4.8cm]{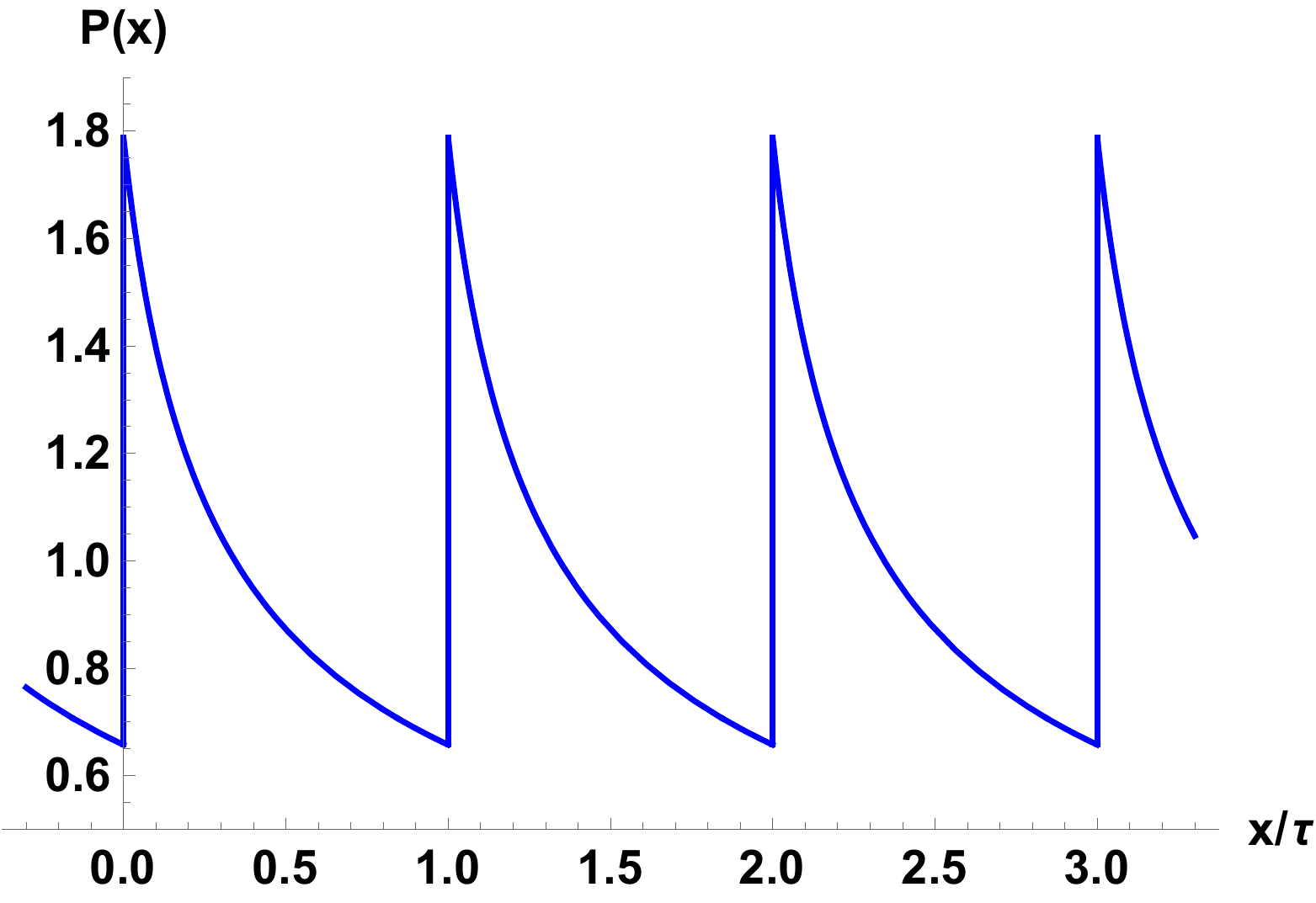}  }
    \hskip0cm
    \subfigure[ ]
    {\includegraphics[width=4.8cm]{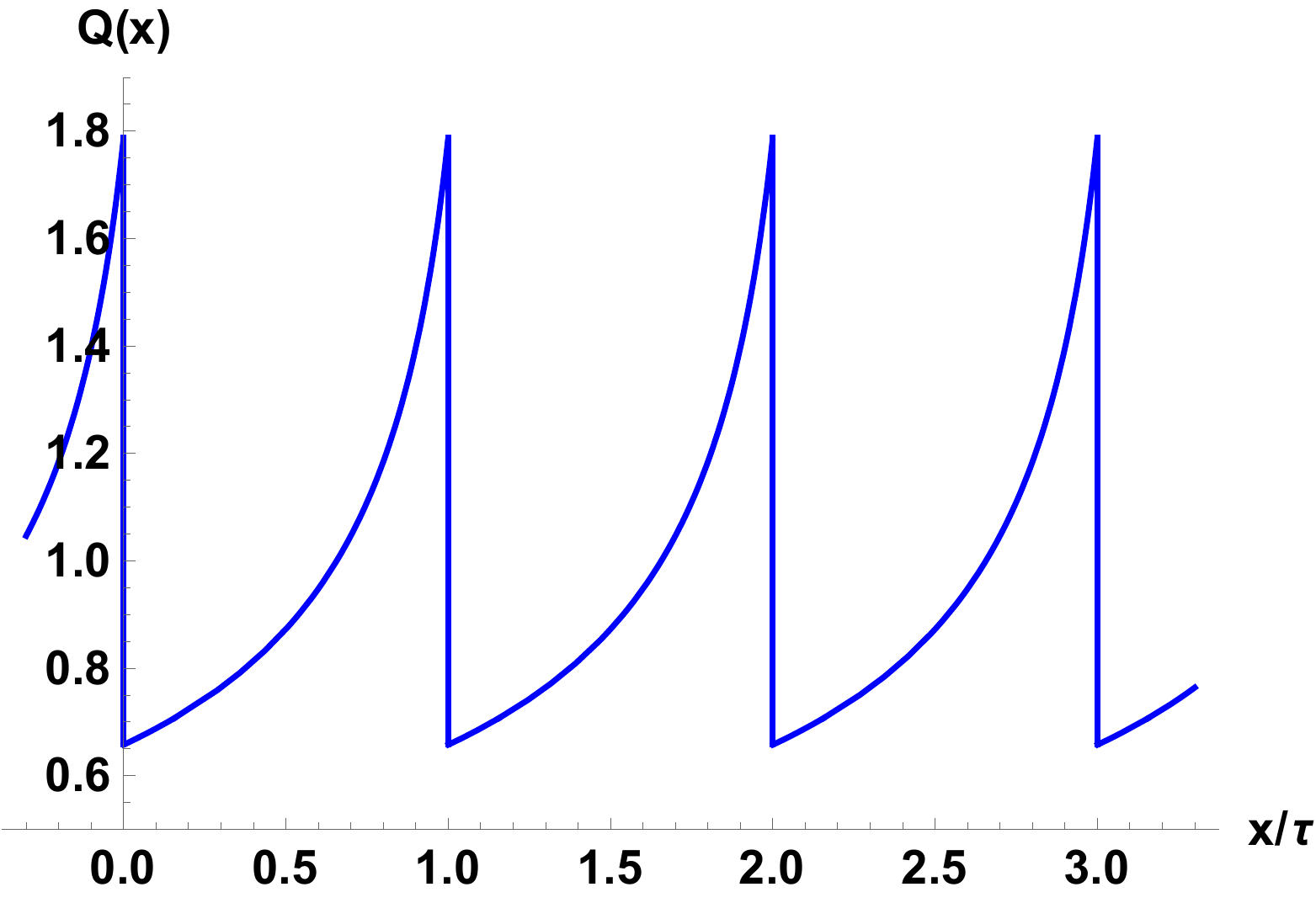}  }
    \hskip0cm

\caption{ Various vacuum configurations of delta function arrays: (a) is a cartoon for the mass-function \eqref{mdelx} and (b) and (c) are the corresponding warping factors to the diagonal matrices in (\ref{PQ_delta_D}). (d) depicts the mass-function (\ref{mGRVV_delta}). In addition (e) and (f) show the warping factors given by (\ref{PQ_delta_GRVV}).}\label{fig:deltaSol}
\end{figure}

As a final interesting example, we consider a mass-function as arrays of delta functions. To find such configurations, we impose a kind of junction condition on field values. This condition is easily obtained by integration of the BPS equations (\ref{D_BPSeq}) and (\ref{D_BPSeq}) near the position of a delta function appearing in the mass-function: $m(x) =\sum_i {q_i}\delta(x-x_i)$. It turns out that the junction condition is 
\begin{align}
\lim_{\epsilon\to 0} \frac{P(x_i+\epsilon)}{P(x_i-\epsilon)} = e^{q_i} ~~,~~\lim_{\epsilon\to 0} \frac{Q(x_i+\epsilon)}{Q(x_i-\epsilon)} = e^{-q_i}.
\end{align}
For the diagonal type solution, one may choose the following mass-function:
\begin{align}\label{mdelx}
m(x)= \sum_{n=-\infty}^{\infty} q_0\left\{ \delta\left(x-n \tau \right) - \delta(x - (n+1/2)\tau) \right\},
\end{align}
where one can notice that the integration of mass-function over one period $\tau$ vanishes. The corresponding solution is the form of $Y^a = P(x) \tilde{Y}_D^a$ and $Y^i = Q(x) \tilde{Y}_D^i$ with
\begin{align}\label{PQ_delta_D}
&P(x) =\left\{ \begin{array}{c} 1\\e^{-q_0}  
\end{array} \right.  ~~
\begin{array}{cc} &n\tau < x <\tau(n + 1/2)  \\  &\tau(n + 1/2) < x <\tau(n + 1)
\end{array},\nonumber\\
&Q(x) =\left\{ \begin{array}{c} e^{-q_0}\\1  
\end{array} \right.  ~~
\begin{array}{cc} &n\tau < x <\tau(n + 1/2)  \\  &\tau(n + 1/2) < x <\tau(n + 1)
\end{array}.
\end{align} 
On the other hand, if we take into account the following mass configuration,
\begin{align}\label{mGRVV_delta}
m(x)=  \sum_{n=-\infty}^{\infty} q_0 \, \delta\left(x-n \tau \right),   
\end{align}  
the GRVV type solution is allowed for this mass configuration with
\begin{align}\label{PQ_delta_GRVV}
&P(x) = \frac{1}{\sqrt{\frac{2 \tau }{e^{2 \text{q0}}-1}+2 (x-n\tau)}} ~~~~(n \tau<x<(n+1)\tau),\nonumber\\
&Q(x) = \frac{1}{\sqrt{\frac{2 e^{2 \text{q0}} \tau }{e^{2 \text{q0}}-1}-2 (x-n\tau)}}~~~~(n \tau<x<(n+1)\tau),
\end{align}
where $q_0=\int_{-\tau/2}^{\tau/2} m(x')dx'$.
We plot the solutions in Figure \ref{fig:deltaSol}. It would be interesting to study physical meaning of these solutions.

\end{document}